\documentclass[twocolumn,showpacs,preprintnumbers,amsmath,amssymb]{revtex4}

\usepackage{graphicx}
\usepackage{epsfig}

\usepackage{longtable}
\usepackage{longtable}

\begin{document}

\title{ Adiabatic fission barriers in superheavy nuclei }

\author{P.~Jachimowicz}
 \affiliation{Institute of Physics,
University of Zielona G\'{o}ra, Szafrana 4a, 65516 Zielona
G\'{o}ra, Poland}

\author{M.~Kowal} \email{m.kowal@fuw.edu.pl}
\affiliation{National Centre for Nuclear Research, Ho\.za 69,
PL-00-681 Warsaw, Poland}

\author{J.~Skalski}
\affiliation{National Centre for Nuclear Research, Ho\.za 69,
PL-00-681 Warsaw, Poland}

\date{\today}

\begin{abstract}

 Using the microscopic-macroscopic model based on the deformed Woods-Saxon
 single-particle potential and the Yukawa-plus-exponential macroscopic
 energy we calculated static fission barriers $B_{f}$ for 1305 heavy and
 superheavy nuclei $98\leq Z \leq 126$, including even - even, odd - even,
 even - odd and odd - odd systems.
 For odd and odd-odd nuclei, adiabatic potential energy surfaces were
 calculated by a minimization over configurations with one blocked neutron
 or/and proton on a level from the 10-th below to the 10-th
 above the Fermi level. The parameters of the model that have been fixed
 previously by a fit to masses of even-even heavy nuclei were kept unchanged.
 A search for saddle points has been performed by the "Imaginary Water Flow"
 method on a basic five-dimensional deformation grid, including triaxiality.
 Two auxiliary grids were used for checking the effects of the mass asymmetry
 and hexadecapole non-axiallity. The ground states were found by energy
 minimization over configurations and deformations.
  We find that the non-axiallity significantly changes first
 and second fission barrier in many nuclei.
  The effect of the mass - asymmetry, known to lower the second, very deformed
  barriers in actinides, in the heaviest nuclei appears
  at the less deformed saddles in more than 100 nuclei.
  It happens for those saddles in which the triaxiallity does not play any
  role, what suggests a decoupling between effects of the mass-asymmetry and
  triaxiality.
  We studied also the influence of the pairing interaction strength on the
  staggering of $B_f$ for odd- and even-particle numbers.
  Finally, we provide a comparison of our results with other theoretical
  fission barrier evaluations and with available experimental estimates.

\end{abstract}


\maketitle

\section{INTRODUCTION}

 Although fission barrier heights $B_{f}$ are not directly measurable
 quantities, i.e. are not quantum observables, they are very useful
 in estimating nuclear fission rates. As the activation energy $E_a$ (per mole)
 in chemistry gives a rate $k$ of a chemical reaction at temperature $T$ via
 the Arrhenius law: $k=Ae^{-E_a/RT}$ ($R$ - the gas constant; $A$ - the
 frequency factor) \cite{tHoff,Arrhe}, the fission barrier gives the fission
 rate $\Gamma_f$ of an excited (as they usually are in nuclear reactions)
 nucleus via:
 $\Gamma_f\sim e^{-B_f/kT_{eff}}$, where $T_{eff}$ is an effective temperature
 derived from the excitation energy, and $k$ - the Boltzman constant.
 For example, knowing fission barriers of possible fusion products
 helps predicting a cross section for a production of a given evaporation
 residue in a heavy ion reaction: one can figure out whether
 neutron or alpha emission wins a competition with fission at each
 stage of the deexcitation of a compound nucleus.
  Moreover, one can try to understand the experimentally established,
 intriguing growth of the total cross sections around Z=118; for its
 correlation with $B_f$, see e.g. Fig. 6 and the related discussion in
 \cite{O1}.
 On the other hand, the prediction of the spontaneous or low energy (i.e. from
 a weakly excited state) fission rates, governed by the regime of the
 collective quantum tunneling, requires an additional knowledge of
  the barrier shape and mass parameters.

 A non-observable status of the fission barrier, again in analogy to that of
 the activation energy in chemistry, is reflected in its possible dependence
 on a reaction type and/or the excitation energy (effective temperature) range.
 This leads to some uncertainty in calculations of fission barriers. In
 particular, it is not clear whether intrinsic configurations should be
 conserved along the level crossings, which increases $B_f$, or the adiabatic
 state should be followed. This is especially relevant for odd-$A$ and
 odd-odd nuclei, in which sharp crossings of levels occupied by the
 odd particle exclude the strictly adiabatic scenario. It is known that if the
 projection of the single-particle angular momentum on the symmetry axis of a
 nucleus $\Omega$ is conserved, the diabatic effect on the fission barrier can
 be huge, see e.g. \cite{Kisomers}. As there is no accepted formula for a
 barrier correction due to the non-adiabaticity, it is usually ignored, even in
  odd-$N$ and/or odd-$Z$ nuclei.
  A general idea is that at the excitation energies close to, and higher than
  the barrier,  but still not inducing sizable dissipative corrections,
  the adiabatic barrier could be used for calculating fission rates.

 Since calculations of potential energy surfaces (PES's) for odd-$A$ and
 odd-odd nuclei involve a repetition of calculations for many low-lying
 quasiparticle states which multiplies the effort (especially in odd-odd
 systems), systematic studies of their fission barriers are rather scarce.
 Up to now, they were provided mainly by the Los Alamos microscopic-macroscopic
 (MM) model and recently by some self-consistent models \cite{Gorbar}.
 The current state of theoretical predictions in fission of even - even nuclei
 (with Z$\geq$ 100) has been discussed recently in \cite{nucl2015}.

 In the present paper we extend our MM model based
 on the deformed Woods-Saxon potential, which up to now was applied mainly
 to even-even nuclei \cite{Kow}, to odd-$A$ and odd-odd SH systems.
 We study a wide range of isotopes which, perhaps, may be of some use for
 astrophysical purposes. The fission barriers are calculated using the
 adiabatic assumption, i.e. they are the smallest possible.
 Since the model has been quite reasonable, in particular in reproducing
 first \cite{Kow} and second \cite{IIbarriers} fission barriers in actinides,
 as well as super- \cite{kowskal} and hyper-deformed \cite{IIIbarriers1,IIIbarriers2} minima, we prefer to keep its
 parameters unchanged. The shell and pairing correction for an odd nucleon
 system is done by blocking the lowest-lying quasiparticle states.
 The modification of the macroscopic energy by including the average
 pairing energy contribution which we introduced for nuclear masses
 in \cite{JachKowSkal2014} is irrelevant for fission barriers.

 The other motivation of our study is to improve the predictions for the
 fission saddles. This requires simultaneously taking into account
 a large number of shape variables \cite{IIbarriers,IIIbarriers2} and relying
 on an {\it in principle} exact method for finding saddles to escape
 errors inherent in the mostly used constrained minimization method, see
  \cite{Moller2009,Dubray}.
 As usual, to make the involved computational effort manageable one has to
 make some compromises which will be discussed in detail.
 The need for a simultaneous consideration of many shape variables in
 PES's calculations is common to all nuclear models, including self-consistent
 theories based on some effective interactions \cite{Schunck}.
 The results on fission saddles obtained up to now in the SH region clearly
 show the great importance of triaxial deformation, neglected in many
  published work.
 A recent study \cite{BroSkal} of barriers within both the MM Woods-Saxon and
 Skyrme SLy6 Hartree-Fock plus BCS models shows that triaxiality
 is even more crucial beyond $Z =$ 126.

  A description of our method of calculations is given in section II. The
  results, details of the additional calculations, and comparisons with
  other calculated barriers are presented and discussed in section III.
  Finally, the conclusions are summarized in section IV.

\section{The Method}

 Multidimensional energy landscapes are calculated within the
 MM model besed on the deformed Woods-Saxon potential \cite{WS}.
The Strutinski shell and pairing correction \cite{STRUT67}
 is taken for the microscopic part.
For the macroscopic part we used the Yukawa plus exponential model \cite{KN}
 with parameters specified in \cite{MUNPATSOB}. Thus, all parameter values
  are kept exactly the same as in all recent applications of the
 model to heavy and superheavy nuclei

 The main point in fission barrier calculations is its reliability which,
 once the model for calculating energy of a nucleus as a function of
 deformation is fixed, hangs on two main
 ingredients: 1) the kind and dimension of the admitted deformation space and
 2) a method applied to the search for saddles.

 Mononuclear shapes can be parameterized via spherical harmonics
 ${\rm Y}_{lm}(\vartheta ,\varphi)$ (for brevity we will just use
 the symbol ${\rm Y}_{\lambda\mu}$ ) by the following equation of the nuclear surface:
    \begin{equation}
   R(\vartheta ,\varphi)= c(\{\beta\}) R_0
 \{ 1+ \sum _{\lambda=1}^{\infty}\sum _{\mu=-\lambda}^{+\lambda} \beta_{\lambda\mu}{\rm Y}_{\lambda\mu}\},
   \label{eq:radius}
\end{equation}
  where $c(\{\beta\})$ is the volume-fixing factor and $R_0$ is the radius of a spherical nucleus.
This parameterization has its limitations; certainly, it is not suitable for
 too elongated shapes.
 However, for moderately deformed saddle points in superheavy nuclei
 it excellently reproduces all shapes generated by other parametrizations,
  e.g. by \cite{Mol2000}, as we checked in numerous tests.

 For nuclear ground states it is possible to confine analysis to
 axially-symmetric shapes, with the expansion truncated at $\beta_{80}$:
   \begin{eqnarray}
 R(\vartheta ,\varphi) &=& c(\{\beta\})R_0 \{ 1 + \beta_{20}  {\rm Y}_{20}+
 \beta_{30}  {\rm Y}_{30}  + \beta_{40}  {\rm Y}_{40}  \nonumber\\
 &+&  \beta_{50} {\rm Y}_{50} + \beta_{60} {\rm Y}_{60} +
 \beta_{70} {\rm Y}_{70} + \beta_{80} {\rm Y}_{80} \}. \nonumber\\
 &&
 \end{eqnarray}
 Thus, a seven dimensional minimization is performed using the gradient method.
 For odd systems, the additional minimization over configurations is
 performed at every step of the gradient procedure. Considered configurations
 consist of the odd particle occupying one of the levels close to the Fermi
 level and the rest of the particles forming a paired BCS state on the
 remaining levels. Ten states above and ten states below the Fermi level have
 been blocked and energy minimized over these configurations.

 The main problem in a search for saddle points is that, since they are neither
 minima nor maxima, one has to know energy on a multidimensional grid of
  deformations (the often used and much simpler method of minimization with
  imposed constraints may produce invalid results
 \cite{Moller2009,Dubray,IIbarriers,Schunck}.
  To find saddles on a grid we used the Imaginary Water Flow (IWF) technique.
 This conceptually simple and at the same time very efficient (from a numerical
  point of view) method was widely used and discussed before
 \cite{Luc91,Mam98,Hayes00,Moeler04,Moller2009,IIbarriers}. The
 number of numerically tractable deformation parameters
 $\{\beta_{\lambda \mu}\}$ is practically limited.
 More than five-dimensional grids, keeping in mind a subsequent interpolation,
 are intractable in calculations for many ($\sim$ 1000) nuclei. Including
 mass- and axially-symmetric deformations ($\beta_{20}$, $\beta_{40}$,
  $\beta_{60}$, $\beta_{80}$ - see \cite{Patyk1,Patyk2,Patyk3,Smol1} together with both,
  mass-asymmetry ($\beta_{30}$, $\beta_{50}$, $\beta_{70}$) and triaxiality
 (at least $\beta_{22}$) would mean at least an eight-dimensional mesh
 and was impossible at present.

 Based on our previous results showing that triaxial saddles are abundant in SH nuclei \cite{Kow}, we consider that quadrupole
 triaxial shapes have to be necessarily included.
 We treated the effects of mass-asymmetry and nonaxial higher multipoles
  as corrections and analysed them at the second stage of calculations.
 A rationale for a lesser importance of mass-asymmetric saddles is that,
 while they constitute a second, more deformed ($\beta_{20}\approx
 0.7 - 0.8)$,
  prominent barrier peak in actinides, their heights are much reduced
  in SH nuclei where they become irrelevant. In the remaining, less deformed
  saddles the mass asymmetry occurs less frequently.
  As to the nonaxial multipoles of higher order, they are less important
  for saddles with small to moderate $\gamma$ [where $\gamma$ is the
  Bohr's quadrupole nonaxiality parameter, cf. Eq. (\ref{betgam})].
  They become important for $\gamma$ closer to $60^o$
  where they are needed to produce oblate shapes having $x$ as the symmetry
  axis.
   Thus, they should be included for nuclei with a large
  oblate g.s. deformation and a short triaxial barrier.
  The additional studies of the mass-asymmetry and higher nonaxial multipoles
   are described in the proper subsections of the Results section.

 Thus, at the first stage, for all 1305 investigated nuclei the saddle points
 were searched in a five dimensional deformation space spanned by:
 $\beta_{20}$, $\beta_{22}$, $\beta_{40}$, $\beta_{60}$, $\beta_{80}$, using
 the IWF technique.
 The appropriate nuclear radius expansion has the form:
  \begin{eqnarray}
 \label{maingrid}
 R(\vartheta ,\varphi) &=& c(\{\beta\})R_0 \{ 1 + \beta_{20} {\rm Y}_{20}
 +\frac{\beta_{22}}{\sqrt{2}}\left[{{\rm Y}_{22} +
 {\rm Y}_{2 - 2}}\right]  \nonumber\\ &+&\beta_{40}  {\rm Y}_{40} +
  \beta_{60} {\rm Y}_{60} + \beta_{80} {\rm Y}_{80}  \}.
\end{eqnarray}
The five-dimensional calculations are  performed on the following deformation
 mesh:
\begin{eqnarray}
\beta_{20} & = &  \ \  0.00 \ (0.05) \ 0.60    \nonumber \\
\beta_{22} & = &  \ \  0.00 \ (0.05) \ 0.45    \nonumber \\
\beta_{40} & = &  \ \  -0.20 \ (0.02) \ 0.20    \nonumber \\
\beta_{60} & = &  \ \  -0.10 \ (0.02) \ 0.10    \nonumber\\
\beta_{80} & = &  \ \  -0.10 \ (0.02) \ 0.10    \nonumber\\
\end{eqnarray}
 This makes a grid of 29250 points which was subsequently interpolated
 to a fivefold denser grid of 50735286 points with the step 0.01 in
 each dimension. On the latter, the saddle point, or rather several saddle
 points - if there were a few of comparable heights within the 0.5 MeV
 energy window - were searched for by means of the IWF procedure.
For odd or odd-odd nuclei, at each grid point we were looking for low-lying
 configurations by blocking particles on levels from the 10-th below to the
 10-th above the Fermi level (in neutrons or/and protons).


  The fact that searches for ground states and for saddles are separated -
   performed using different deformation spaces - allows saving some
  number of deformation parameters in Eq. (\ref{maingrid}).
 This is equivalent to
  assuming that the fission saddles have mostly prolate deformations large
 enough to make nonaxial deformations of multipolarity $\lambda\geq 3$ less
  important. One has to check this assumption afterwards and separately
  treat nuclei in which the inclusion of nonaxial deformations with
  $\lambda\geq 4$ is necessary.

  Although, as mentioned before, in SH nuclei the second barriers at large
  deformations are usually smaller than the first one or do not exist at all,
  for $Z=98$-101 the mesh (4) was extended to $\beta_{20}=1.5$ and the second
  saddles were searched for by the IWF technique. It turned out that these
  more deformed barriers are indeed mostly smaller than the first ones and
  decrease with increasing $Z$.
  Only in Cf isotopes with $N=134-160$ there were some second saddles
  (at $\beta_{20}\approx 0.9$) higher than the first one by at most 0.5 MeV.
  However, even those saddles were lowered by at least 1 MeV after including
  the mass-asymmetry. Therefore, we have reasons to believe that the range of
 $\beta_{20}$ in (4) is sufficient for knowing the height of the fission
  barrier in the whole studied region.

 \section{Results and discussion}

In the present paper we have systematically calculated fission-barrier
heights $B_{f}$ as the energy difference between the saddle point and
 the ground state. The saddle point is defined as the minimum over possible
 fission paths of the maximal energy along the path.
Let us emphasize that the calculations presented here have
been performed without adding any zero-point vibration energy.
 We have included 1305 heavy and superheavy nuclei with
 proton numbers $98\leq Z \leq 126$ and neutron numbers
 in the range $134\leq N \leq 192$, with the smallest $N$ for a given $Z$
 increasing by one with every step in $Z$. All obtained barriers have been
 collected in Table \ref{bartot}.
 On all PES's presented here, energy is normalized in such a way that its
  macroscopic part is set to zero at the spherical shape.

%







 \subsection{Potential Energy surfaces}

\begin{figure}
 \includegraphics[width=1.2\linewidth,height=3.2in]{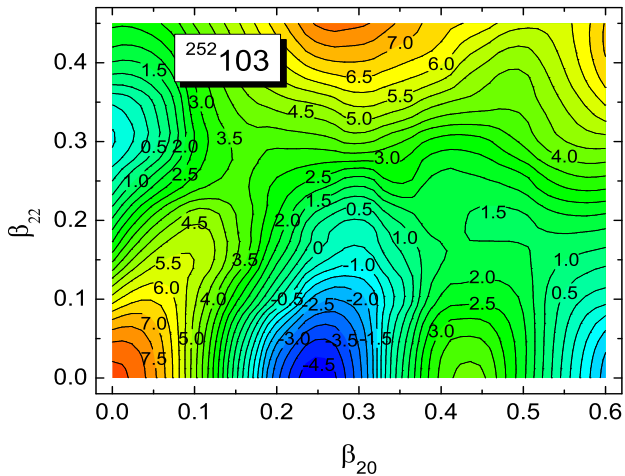}
\caption{Energy surface, $E-E_{mac}(sphere)$, for $Z=103$ and $N=149$.}
 \label{figb103_149}.
\end{figure}

\begin{figure}
\includegraphics[width=1.2\linewidth,height=3.2in]{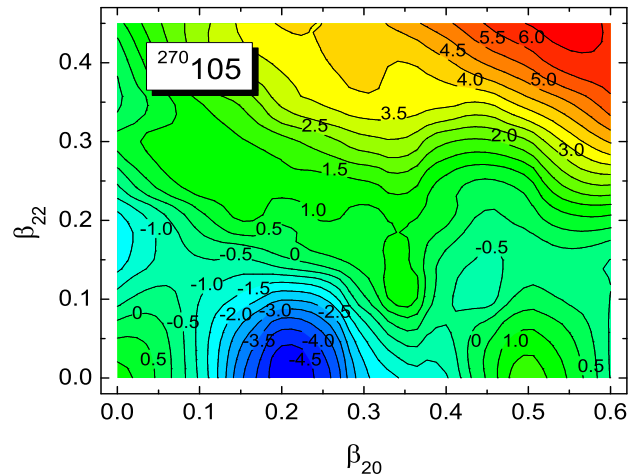}
\caption{The same as in \ref{figb103_149} but for $Z=105$ and $N=165$.}
 \label{figb105_165}.
\end{figure}

\begin{figure}
 \includegraphics[width=1.2\linewidth,height=3.2in]{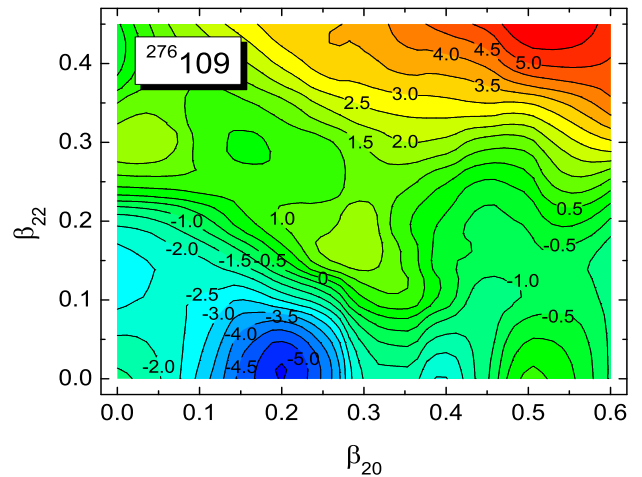}
\caption{The same as in \ref{figb103_149} but for $Z=109$ and $N=167$.}
 \label{figb109_167}.
\end{figure}

\begin{figure}
 \includegraphics[width=1.2\linewidth,height=3.2in]{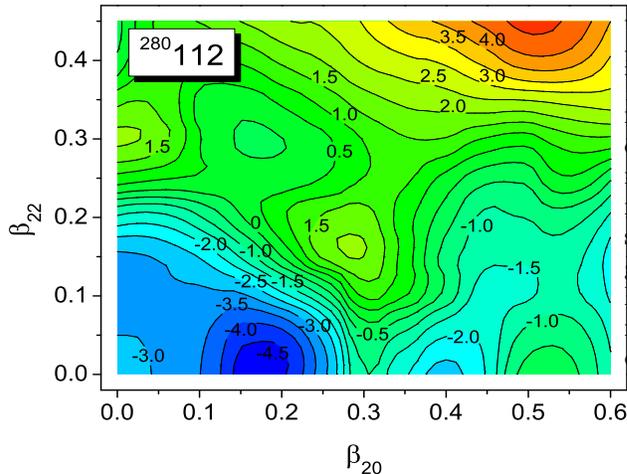}
 \caption{The same as in \ref{figb103_149} but for $Z=112$ and $N=168$.}
 \label{figb112_168}.
\end{figure}

\begin{figure}
 \includegraphics[width=1.2\linewidth,height=3.2in]{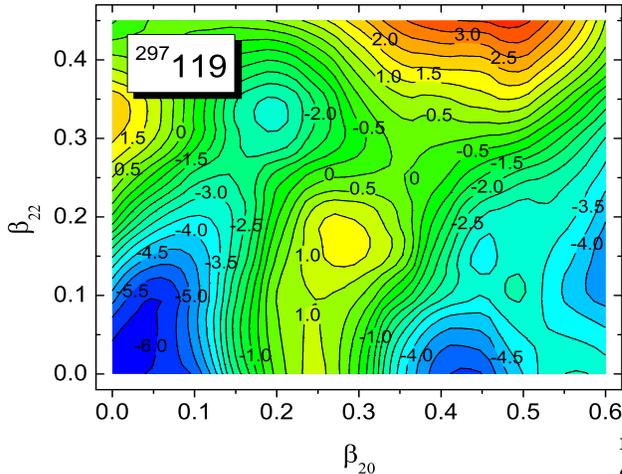}
 \caption{The same as in \ref{figb103_149} but for $Z=119$ and $N=178$.}
 \label{figb119_178}.
\end{figure}

 Some idea about the positions of ground states, secondary minima and saddles
 may be gained from PES's. Chosen examples are shown in figures for:
 $^{252}$Lr - Fig. \ref{figb103_149},
 $^{270}$Db - Fig. \ref{figb105_165}, $^{276}$Mt - Fig. \ref{figb109_167},
 $^{280}$Cn - Fig. \ref{figb112_168}, and $^{297}$119 - Fig. \ref{figb119_178}.
 Overall evolution of ground states with increasing $Z$ from prolate to
 spherical can be seen there. In some nuclei one can see multiple saddles
 of which the one defining the fission barrier should be properly chosen.
 Sometimes the saddles between competing minima can be important,
 therefore the determination of all saddles on the map is necessarily needed.

 The energy landscapes Fig. 1-5 were obtained by minimizing energy on the
  5D grid (\ref{maingrid}) with respect to $\beta_{40}$, $\beta_{60}$ and
  $\beta_{80}$. One should be aware of two related circumstances: 1)
  As the grid Eq. (\ref{maingrid}) does not include nonaxial deformations
  $\lambda\geq 4$, the axial deformations $\lambda= 4, 6, 8$ {\it with respect
  to the $x$-axis} cannot be reproduced, so the landscapes are inexact
  around the oblate $\gamma=60^o$ axis. 2)
  A reduction of a $n$-dimensional grid of energy values via the minimization
 over $n-2$ deformations sometimes leads to an energy surface composed from
  disconnected patches, corresponding to multiple minima in the auxiliary
 (those minimized over) dimensions. This can distort the picture of the
   barrier (actually, a reduction of multi-dimensional data to a
  two-dimensional map is a general problem).

 With these reservations in mind, one can still explore some of the details
 shown in the maps. In particular, the prolate g.s. minimum with
 strongly nonaxial first saddle point at $\beta_{20} =0.41$ and
 $\beta_{22} =0.18$ is visible in $^{252}$Lr. One can notice that the
 axially symmetric saddle lies more than 2 MeV higher.
A slightly less steep, prolate g.s. minimum and a gently emerging
 second minimum is visible in Fig \ref{figb105_165} for $^{270}$Db.
 The triaxial saddle at $\beta_{20} =0.52$ and $\beta_{22} =0.13$ has
 a smaller triaxiality $\gamma$ than the saddle in $^{252}$Lr.
 A decrease in barrier height due to triaxiality is $\approx$ 2 MeV,
 Fig. \ref{figb105_165}.

  In a heavier nucleus $^{276}$Mt, a prolate deformation of the g.s. is clearly
  smaller than in $^{252}$Lr, see Fig \ref{figb109_167}. The second minimum,
  which was barely outlined in $^{270}$Db, is more pronounced here,
  giving the fission barrier a double-hump structure.
  The deformation $\beta_{20}\approx 0.5$ of the second saddle is
  much smaller than that of the second barriers in actinides.
  Thus, a two-peak structure of the barrier in SH nuclei may be viewed as
  a result of a division (split) of the first barrier, occurring with growing
  $Z$. The higher second axial saddle is lowered by triaxiality
   by $\approx$ 1.5 MeV, but eventually is still higher than the first axial
  saddle.

For $^{280}$Ds a topology of the PES is even more complicated.
We see several minima: prolate - the g.s. and a superdeformed one, and a
 shallow oblate. The map shows also a few saddles. The axially deformed saddle
 point at $\beta_{20} =0.3$ has a similar height as the nonaxial saddle at
 $\beta_{20} =0.54$ and $\beta_{22} =0.12$. It follows from the IWF calculation
 that the second fission barrier is nonaxial in this case. The axial
  second saddle is lowered by $\approx 1$ MeV owing to the nonaxiallity.

The nucleus $Z=119$, $N=178$ is spherical in its g.s. - Fig. \ref{figb119_178}.
 There is a secondary oblate minimum (whose depth is underestimated in the map
 due to omission of nonaxial $\lambda=4, 6$ deformations). There is a low
 triaxial second saddle at $\beta_{20}\approx 0.5$ and two "first saddles"
 with different triaxiality, of which the one with a larger $\gamma$ is the
 fission saddle.

 Still another type of PES, typical of nuclei with the superdeformed oblate
 g.s., is presented in Fig.\ref{figb4} in the subsection C.

 \subsection{Role of the mass asymmetry}

 To study the effect of the reflection (mass) - asymmetry on the
 fission barriers, a two-step procedure has been performed.
 At the first stage, we have checked the stability of all the saddles found
 on the basic 5D mesh (the first, the second, ..., axially symmetric or
 triaxial, of energy within 0.5 MeV of the highest saddle) against the
  mass-asymmetry.  This was done by a 3D energy minimization with respect to
  $\beta_{30}$, $\beta_{50}$ and $\beta_{70}$ around each saddle. Since most
 of the saddles are non-axial, the most general version of our Woods-Saxon
 code had to be used. In this case, when both symmetries (axial and mass
 symmetry) are broken simultaneously, the nuclear shapes are defined by the
  following equation of the nuclear surface:
   \begin{eqnarray}
 R(\vartheta ,\varphi)= R_0 c(\{\beta\})\left\{ 1 \right.&+&  \beta_{20}
 {\rm Y}_{20}+\frac{\beta_{22}}{\sqrt{2}} \left[{{\rm Y}_{22} + {\rm Y}_{2 - 2}}\right]  \nonumber\\
 &+& \beta_{30}  {\rm Y}_{30} + \beta_{40}  {\rm Y}_{40} + \beta_{50}  {\rm Y}_{50}  \nonumber\\
 &+& \beta_{60} {\rm Y}_{60} +  \beta_{70}  {\rm Y}_{70} + \beta_{80} {\rm Y}_{80}  \}.
\end{eqnarray}

 It turned out that this minimization lowers energy of only those saddles in
  which: i) there is no triaxiality, ii) deformation $\beta_{20}\approx 0.3$.
 This supports an often expressed conventional "wisdom", that the
 mass-asymmetry and triaxiality effects on fission saddle are decoupled.
 This is why, at the second step of the procedure, we could carry out a full
 IWF analysis on a grid including only axially-symmetric deformations:
 $\beta_{20},\beta_{30},\beta_{40},\beta_{50},\beta_{60},\beta_{70},\beta_{80}$,
  with $\beta_{20}$ restricted to a quite short interval $0.25 - 0.40$:

 \begin{eqnarray}
\beta_{20} & = &  \ \   0.25 \ (0.05) \ 0.40    \nonumber \\
\beta_{30} & = &  \ \   0.00 \ (0.05) \ 0.25    \nonumber \\
\beta_{40} & = &  \ \  -0.15 \ (0.05) \ 0.20    \nonumber \\
\beta_{50} & = &  \ \   0.00 \ (0.05) \ 0.15    \nonumber\\
\beta_{60} & = &  \ \  -0.10 \ (0.05) \ 0.10    \nonumber\\
\beta_{70} & = &  \ \   0.00 \ (0.05) \ 0.15    \nonumber\\
\beta_{80} & = &  \ \  -0.10 \ (0.05) \ 0.10.    \nonumber\\
\end{eqnarray}

 This seven-dimensional grid, composed of 76800 deformations,
  was subject to the fivefold interpolation in all directions before it was
 used in the IWF procedure. This means that the IWF calculations have been
 performed on the grid containing 1 690 730 496(!) points.
 We have made such 7-dimensional analysis for more than 100 nuclei, for which
 the effect of minimization was greater than 300 keV. Results for these nuclei
 are shown
 in Table \ref{Octu}. The rest of 127 cases shown in Table \ref{Octu} are the
 test nuclei, in which the effect of the minimization was smaller than 0.3 MeV.
 The results for these additional nuclei allow to appreciate whether the
  (in principle exact) IWF method could produce a greater effect that the
  (inexact) minimization.

\begin{figure}[h!]
 \includegraphics[width=1.3\linewidth,height=5.0in]{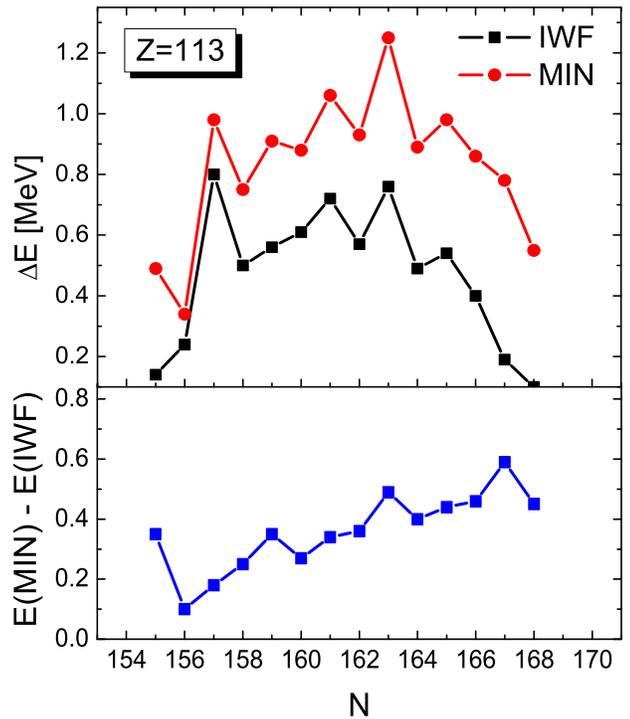}
 \caption{ Top panel:
 The fission barrier lowering by the mass-asymmetry obtained by the
 (in principle exact) Imaginary Water Flow method - IWF and by the
 (easier, but sometimes misleading) minimization method - MIN.
  Bottom panel: The difference between both methods in MeV (in principle -
  the error in the barrier height due to the minimization method). }
 \label{oct}
\end{figure}

 As one can see, the adopted procedure allowed to omit the problem of
 searching for a saddle by using the (inexact) minimization method which
 is not always reliable \cite{IIbarriers,Moller2009}. For example, for Z=118
 and N=165, the discussed effect resulting from the minimization
 amounts to 0.44 MeV, which, just in this case, is quite similar to 0.46 MeV
 obtained from the IWF technique; however, in Z=113 and N=163 one obtains
  $\approx 0.5$ MeV difference between saddles obtained by both methods.
 In this particular nucleus, the $\approx 0.77$ MeV barrier
 lowering by the mass-asymmetry is the largest among all studied nuclei.
 It should be also noted that for the isotopes of Z = 113 the
 effect of the mass-asymmetry is particularly large, see the the top panel
 in Fig. \ref{oct}.

 In the bottom panel of Fig. \ref{oct}, we show the difference between the
 results of the both methods - the minimization - (MIN) and
 "Imaginary Water Flow" - (IWF).
 One can see that this difference increases with the neutron number.
 In particular,
 there is practically no effect derived from the mass-asymmetry in $^{281}113$
 when IWF is used. On the contrary, the approach based on minimization
 suggests still a quite substantial (spurious) effect (0.55 MeV).
 One might notice that our conclusion concerning decoupling of the variables
 describing the axial and reflection asymmetries is in a delicate
 contradiction with the studies \cite{Lu}.

\begingroup



\begin{table*}

\caption{\label{Octu} Mass(reflection)-asymmetry effect on the fission barrier
 from the minimization - MIN and from the Imaginary Water Flow method - IWF
 (in MeV).}

 \begin{ruledtabular}

\begin{tabular}{|cccccccccc|}

&     N &   IWF     &  MIN      &  N   &   IWF     &  MIN     &    N  &   IWF      &  MIN   \\

\noalign{\smallskip}\hline\noalign{\smallskip}
&		&	$\mathbf{Z=109}$	&		&		&	$\mathbf{Z=114}$	&		&		&	$\mathbf{Z=117}$  & \\		
&	157	&	0.39	&	0.81	&	155	&	0.28	&	0.59	&	157	&	0.24	&	0.34	\\		
&	158	&	0.22	&	0.42	&	156	&	0.14	&	$<$0.30	&	158	&	0.28	&	$<$0.30	\\		
&	159	&	0.54	&	0.45	&	157	&	0.72	&	0.83	&	159	&	0.24	&	0.34	\\		
&	160	&	0.31	&	0.54	&	158	&	0.46	&	0.46	&	160	&	0.12	&	$<$0.30	\\		
&		&	$\mathbf{Z=110}$&   &	159	&	0.67	&	0.68	&	161	&	0.26	&	$<$0.30	\\		
&	157	&	0.41	&	0.69	&	160	&	0.45	&	0.66	&	165	&	0.36	    &	0.39	\\		
&	158	&	0.19	&	0.31	&	161	&	0.53	&	0.79	&	166	&	0.23	&	$<$0.30	\\		
&	159	&	0.52	&	0.46	&	162	&	0.42	&	0.64	&	167	&	0.19	&	0.50	\\		
&	160	&	0.50	&	0.40	&	163	&	0.58	&	0.65	&	168	&	0.07	&	$<$0.30	\\		
&	161	&	0.43	&	0.47	&	164	&	0.40	&	0.63	&	169	&	0.05	&	0.37	\\		
&	162	&	0.35	&	0.31	&	165	&	0.42	&	0.65	&		&	$\mathbf{Z=118}$	&		\\		
&		&	$\mathbf{Z=111}$	&&	166	&	0.38	&	0.53	&	163	&	0.30	&	0.32	\\		
&	157	&	0.49	&	0.97	&	167	&	0.11	&	0.68	&	164	&	0.23	&	$<$0.30	\\		
&	158	&	0.36	&	0.78	&	168	&	0.06	&	0.41	&	165	&	0.46	&	0.44	\\		
&	159	&	0.61	&	0.83	&		&	$\mathbf{Z=115}$	&&	166	&	0.28	&	0.31	\\		
&	160	&	0.67	&	0.85	&	157	&	0.28	&	0.64	&	167	&	0.20	&	0.63	\\		
&	161	&	0.87	&	0.89	&	158	&	0.25	&	0.50	&	168	&	0.15	&	0.39	\\		
&	162	&	0.66	&	0.80	&	159	&	0.34	&	0.49	&		&	$\mathbf{Z=119}$	&		\\		
&	163	&	0.56	&	0.83	&	160	&	0.39	&	0.38	&	165	&	0.46	&	0.57	\\		
&	164	&	0.58	&	0.68	&	161	&	0.56	&	0.58	&	166	&	0.33	&	0.37	\\		
&	166	&	0.48	&	0.49	&	162	&	0.42	&	0.39	&	167	&	0.34	&	0.49	\\		
&		&	$\mathbf{Z=112}$	&&	163	&	0.46	&	0.54	&	168	&	0.27	&	0.32	\\		
&	157	&	0.57	&	0.83	&	164	&	0.49	&	0.45	&	169	&	0.31	&	0.57	\\		
&	158	&	0.32	&	0.45	&	165	&	0.47	&	0.60	&	170	&	0.24	&	0.38	\\		
&	159	&	0.58	&	0.55	&	166	&	0.53	&	0.54	&	171	&	0.23	&	0.32	\\		
&	160	&	0.60	&	0.49	&	167	&	0.42	&	0.80	&		&$\mathbf{Z=120}$		&		\\		
&	161	&	0.51	&	0.60	&	168	&	0.20	&	0.55	&	165	&	0.39	&	0.38	\\		
&	162	&	0.53	&	0.48	&	169	&	0.13	&	0.31	&	166	&	0.17	&	$<$0.30	\\		
&	163	&	0.56	&	0.64	&	170	&	0.07	&	0.30	&	167	&	0.20	&	0.49	\\		
&	164	&	0.44	&	0.43	&		&	$\mathbf{Z=116}$	&&	168	&	0.15	&	$<$0.30	\\		
&	165	&	0.33	&	0.48	&	155	&	0.40	&	0.41	&	169	&	0.10	&	0.46	\\		
&	166	&	0.34	&	0.34	&	156	&	0.19	&	$<$0.30	&		&	$\mathbf{Z=121}$	&		\\		
&	167	&	0.20	&	0.35	&	157	&	0.36	&	0.52	&	165	&	0.25	&	0.40	\\		
&		&	$\mathbf{Z=113}$	&&	158	&	0.26	&	0.34	&	166	&	0.23	&	$<$0.30	\\		
&	155	&	0.14	&	0.49	&	159	&	0.35	&	0.44	&	167	&	0.38	&	0.52	\\		
&	156	&	0.24	&	0.34	&	160	&	0.28	&	0.49	&	168	&	0.31	&	0.34	\\		
&	157	&	0.80	&	0.98	&	161	&	0.40	&	0.44	&	169	&	0.36	&	0.60	\\		
&	158	&	0.50	&	0.75	&	162	&	0.33	&	0.37	&	170	&	0.30	&	0.43	\\		
&	159	&	0.56	&	0.91	&	163	&	0.48	&	0.54	&		&	$\mathbf{Z=122}$	&		\\		
&	160	&	0.61	&	0.88	&	164	&	0.40	&	0.38	&	164	&	0.00	&	$<$0.30	\\		
&	161	&	0.72	&	1.06	&	165	&	0.46	&	0.50	&	165	&	0.21	&	$<$0.30	\\		
&	162	&	0.57	&	0.93	&	166	&	0.33	&	0.40	&	166	&	0.12	&	$<$0.30	\\		
&	163	&	0.76	&	1.25	&	167	&	0.30	&	0.38	&	167	&	0.19	&	0.31	\\		
&	164	&	0.49	&	0.89	&	168	&	0.11	&	$<$0.30	&	168	&	0.11	&	$<$0.30	\\		
&	165	&	0.54	&	0.98	&	169	&	0.09	&	0.32	&	169	&	0.10	&	0.45	\\		
&	166	&	0.40	&	0.86	&		&    		&		    &		&	$\mathbf{Z=123}$	    &		\\		
&	167	&	0.19	&	0.78	&		&		    &		    &	166	&	0.06	&	$<$0.30	\\		
&	168	&	0.10	&	0.55	&		&		    &    		&	167	&	0.08	&	0.35	\\		
&		&		&		&		&		&		    &		    &$\mathbf{Z=124}$		&		\\		
&		&		&		&		&		&		    &	165	&	0.23	&	0.31	\\		
&		&		&		&		&		&		    &	166	&	0.06	&	$<$0.30	\\		
&		&		&		&		&		&	   	    &	167	&	0.10	&	0.32	\\		
&		&		&		&		&		&		    &		&		    &				\\
																																						
\noalign{\smallskip}

\end{tabular}

\end{ruledtabular}

\end{table*}


\endgroup

 \subsection{Role of the triaxiality}

 The importance of including triaxiality in a calculation of fission barrier
 heights was indicated many times before \cite{CWIOK92,CWIOK94,CWIOK96,GHERGH99,DUTTA00,DECHARGE2003,BONNEAU04,Cwiok05,DOB2007,KOWAL2009}.
 In particular, it was shown that the effect of {\it both}
 quadrupole and a general hexadecapole nonaxiality, when accounted for
 within the {\it nonexact} method of constrained minimization (used generally
 in all selfconsistent studies), may reach 2.5 MeV
 for some superheavy even-even nuclei, see Fig. 5 in \cite{Kow}.
 Here, we extend our previous discussion of its role to
 the odd and odd-odd nuclei and, at the same time, improve the treatment
  by employing the exact IWF method in potentially most interesting cases.

\begin{figure}[h!]
 \includegraphics[width=1.2\linewidth,height=3.5in]{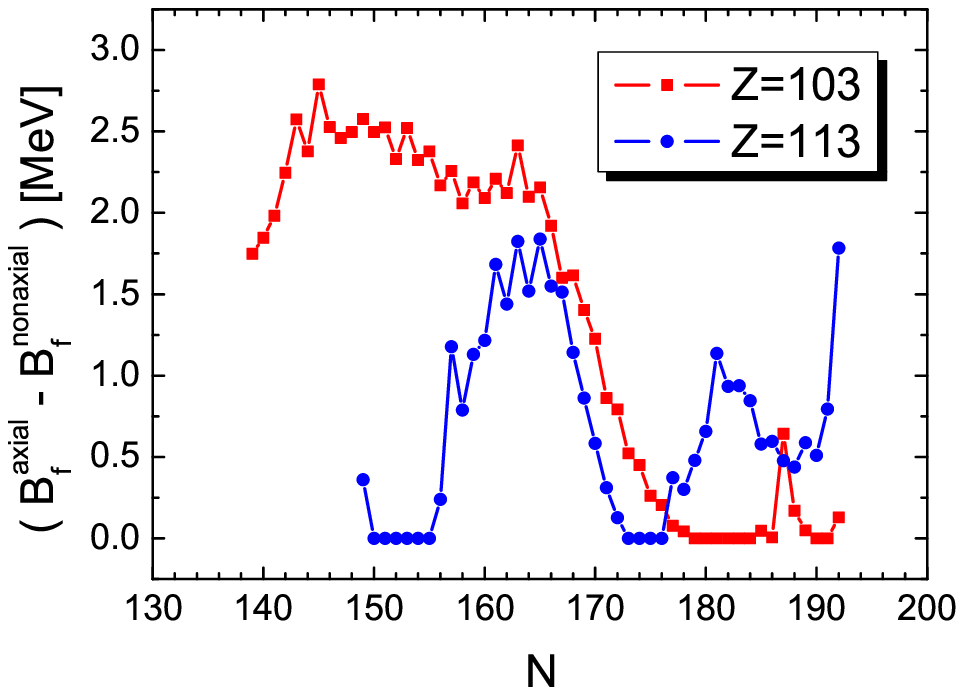}
 \caption{ Effect of the non-axiallity on the fission barrier heights (see
  text for further explanations).}
 \label{nonaxial}
\end{figure}

 By using the original 5D mesh (4) we have obtained saddles with
 {\it quadrupole} nonaxiality for about 900 nuclei, what constitutes more than
 70 \% of all fission barriers. We illustrate this conspicuous effect in Fig.
 \ref{nonaxial} on the example of two isotopic chains, Z=103 and 113.

 We show the difference between axial and nonaxial barriers in these nuclei.
 One can see that for lighter Lawrencium isotopes the effect of nonaxiality is
 quite considerable. Starting with $N=$164, it is weakening quickly and finally
 vanishes for $N\geq$ 176. Somewhat different dependence of the effect on the
  neutron number occurs in $Z=113$ isotopes. The maximum lowering of the barrier
  of more than 1.5 MeV occurs for $N\approx 165$, there is a second maximum at
 $N=179$, and the effect becomes large again at $N=192$. Inbetween, for
  $N\approx 154$ and $N\approx 174$, there is no effect at all.
  Thus, the effect of nonaxiallity has to be studied carefully, indeed.

 Another task is to consider the influence of the hexadecapole nonaxiality,
 namely: $\beta_{42}, \beta_{44}$ in Eq. \ref{eq:radius}, on the fission
 barriers. The unconstrained inclusion of these shapes would lead to a 7D grid
 which is too much for now.
  To evaluate the effect without increasing the grid
 dimension we constrained $\beta_{42}$ and $\beta_{44}$ to be
 functions of the quadrupole nonaxial deformation $\beta_{22}$, or actually
 $\gamma$, and $\beta_{40}$, in a well known manner \cite{geom}.
 Using the conventional notation:
\begin{eqnarray}
 \label{betgam}
\beta& = &\sqrt{\beta_{20}^2+\beta_{22}^2} , \nonumber\\
 \gamma& = & {\rm arctg}\frac{\beta_{22}}{\beta_{20}},
\end{eqnarray}
 the following form of Eq. \ref{eq:radius} was used:

  \begin{eqnarray}
  \label{b4}
 R(\vartheta ,\varphi)= c({\beta})R_0 \left\{ 1 \right.&+&  \beta\cos{(\gamma)}  {\rm Y}_{20} \nonumber\\
 &+& \frac{\beta \sin{(\gamma)}}{\sqrt{2}}\left[{{\rm Y}_{22} + {\rm Y}_{2 - 2}}\right]  \nonumber\\
 &+&\beta_{40}  \frac{1}{6}(5 \cos^{2}{(\gamma)} +1 ){\rm Y}_{40} \nonumber\\
 &-&\beta_{40}  \frac{1}{6}\sqrt{\frac{15}{2}} \sin{(2 \gamma)} \left[{{\rm Y}_{42} + {\rm Y}_{4 - 2}}\right] \nonumber\\
 &+&\beta_{40}  \frac{1}{6}\sqrt{\frac{35}{2}} \sin^{2}{(\gamma)} \left[{{\rm Y}_{44} + {\rm Y}_{4 - 4}}\right] \nonumber\\
 &+& \beta_{60}  {\rm Y}_{60} + \beta_{80}  {\rm Y}_{80}  \}.
\end{eqnarray}

  On this 5D grid, the hexadecapole nonaxiality (but not the $\beta_{60}$ and
  $\beta_{80}$ terms) preserves the modulo-60$^o$
  invariance in $\gamma$, so, in particular, the parameter $\beta_{40}$
  describes a deformation which is axially symmetric around the $z$ axis
  at $\gamma=0^o$ and around the $x$ axis at $\gamma=60^o$, which allows
  to better approximate energy at oblate shapes. For this reason, while the
  original mesh Eq. (\ref{maingrid}) may be expected more reliable
 for barriers at
  small $\gamma$, the one of Eq. (\ref{b4}) is better for saddles closer to
  $\gamma=60^o$, like those in nuclei with well- or super-deformed oblate
 ground states.

 Our method of proceeding is analogous to that used in the study of the
 mass-asymmetry. The difference is that we do not have to perform the
 first step: a minimization with respect to $\beta_{42}$ and $\beta_{44}$
 at the saddles found from the grid Eq. (\ref{maingrid}).
 Such calculations were already
  done in the previous studies of the effect of nonaxial deformations of higher
 multipolarity on the fission barrier in heaviest nuclei
 \cite{KowSob1,KowSob2,KowSob3,SobJachKow}. We know that the minimization
  gave the largest effect in the following four regions of nuclei,
  see Fig. 2 in \cite{SobJachKow}: (I) $Z \approx 122$, $N \approx 160$ -
 up to 1.5 MeV, and a $\sim$ 3 times smaller effect for nuclei with larger
 $N$ and $Z>120$, (II) $Z \approx 110$, $N \approx 146$ - up to 1 MeV,
 (III) $Z\approx114$, $N\approx184$ - up to 1 MeV, and (IV)
 $Z\approx104$, $N\approx170$ - up to 0.4 MeV.

  By applying the IWF method on the mesh Eq. (\ref{b4}) we have found
 the saddles for a dozen of nuclei from the last three regions,
 for which the effect of
  minimization was the largest. It turned out that, compared to saddles
  found on the original grid Eq. (\ref{maingrid}), they were lowered by
  less than 150 keV in the region (II), by less then 100 keV in the region
  (III), and even increased by $\sim$ 100 keV in the region (IV).
 On this basis we conclude that the lowering of the fission saddles
  found by the minimization in \cite{Kow,KowSob3} in these three regions
  is in a large measure a spurious effect which mostly vanishes when saddles
  are fixed by a proper method.

  On the contrary, the substantial effect (up to $\approx 1$ MeV) of the
  nonaxial hexadecapole in the region (I), although smaller than found by
  the minimization, survives in the exact IWF treatment. This might be expected
  as these are very heavy $Z\geq 119$ nuclei with short barriers and oblate
  (also superdeformed) ground states, so $\beta_{42}$ and $\beta_{44}$
   are necessary to reproduce energy in the vicinity of the oblate axis.
   Therefore, in the whole region of nuclei with $Z\geq 118$ we calculated
  triaxial barriers by the IWF method using the mesh Eq. (\ref{b4}) and then
  selected the proper fission barriers from two 5D calculations.

\begin{figure}[h!]
 \includegraphics[width=1.2\linewidth,height=3.2in]{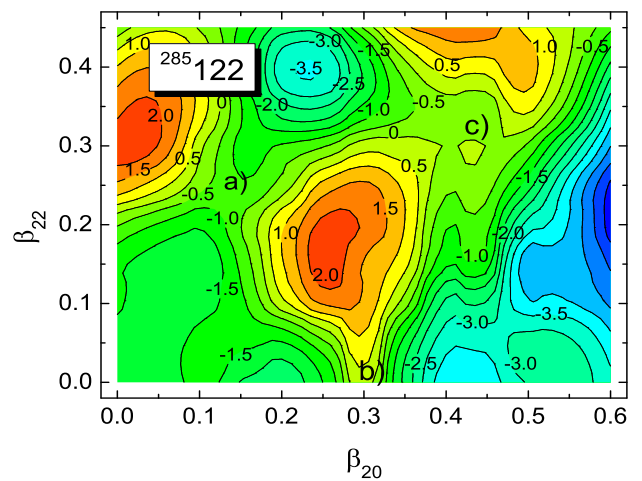}
 \caption{Energy surface, $E-E_{mac}(sphere)$, for the nucleus $Z=122$,
 $N=163$, resulting from the calculation according to Eq. (\ref{b4}).}
 \label{figb4}
\end{figure}

  Three types of saddles in nuclei from the region (I) are shown
 for a very heavy and exotic nucleus $^{285}122$ in Fig. \ref{figb4}.
 The landscape was created from the 5D mesh Eq. (\ref{b4}).

  This nucleus has a global superdeformed oblate (SDO) minimum with
 the quadrupole deformation $\beta_{20} = -0.455$ (spheroid with the axis ratio $\approx$ 3:2).
 It represents a neutron-deficient area of superheavy nuclei  according to
 recent predictions \cite{SDO}. These intriguing SDO minima were already
 confirmed, as the global ones, by various self-consistent models
 \cite{staszproch,skal2015}.
  There is a saddle close to the oblate axis, separating the SDO g.s.
 from the wide minimum near the spherical shape - type a);
  the axially symmetric saddle is designated as b).
  One fission path may go through the saddles a) and b), the higher of which
 would define the barrier along this path. The second fission path goes through
 a triaxial saddle of type c) at $\beta_{20}\approx 0.4$, $\gamma\approx 35^o$.
 The fission barrier of $B_{f}=3.6$ MeV corresponds to the saddle c) as found
 by using the grid Eq. (\ref{b4}). It turns out that saddles of type a) and c)
 are much lowered by including $\beta_{42}$, $\beta_{44}$, the first usually
 more than the second.

  Table \ref{hexeffect} summarizes the effect of nonaxial hexadecapole
 on the barriers in the region (I). It contains 75 nuclei in which the barrier
  lowering is greater than 300 keV.
  The most frequent saddle type in the region (I), on both grids, is c), but
  there are also more complicated cases in which the saddle type changes when
 $\beta_{42}$ and $\beta_{44}$ are included. The largest effect of 1.167 MeV
  occurs in the nucleus $Z=125$, $N=163$.


 Let us remark that the difference between the results of the constrained
 minimization and the IWF method for the nonaxial hexadecapole is the main
 source of the discrepancy between the current fission barriers and those
 published in \cite{Kow} for even-even nuclei.

\begingroup



\begin{table*}

\caption{\label{hexeffect} The barrier lowering (in MeV) greater than 0.3 MeV
 in nuclei $Z\geq 118$, in particular in those with SDO ground states,
 from the IWF calculations on the 5D mesh including $\beta_{42}$ and
 $\beta_{44}$ according to Eq. (\ref{b4}). Also reported is the associated
 change in the saddle type (for a description of saddle types see text);
 no entry means that a c-type saddle results from both grids,
 Eq. (\ref{maingrid}) and (\ref{b4}).}

 \begin{ruledtabular}

\begin{tabular}{|cccccccccc|}

&     N &  $\Delta B_f$   &  saddle   &  N   &   $\Delta B_f$     &  saddle  &    N  &   $\Delta B_f$   &  saddle   \\

\noalign{\smallskip}\hline\noalign{\smallskip}
&		&	$\mathbf{Z=119}$	&		&		&	$\mathbf{Z=122}$	&		&		&	$\mathbf{Z=125}$	\\	
&	155	&      0.597 &                  &	158	&	0.779	&	               	&	161	&	1.083&	               	\\
&	156	&	0.482   &                & 159	&	0.959	&	               	&	162	&	0.958	&                \\
&	157	&	0.472	&               &	160	&	0.807	&               &	163	&	1.167	&                \\
&	158	&	0.566   &               &	161	&	0.731
&               &	164	&	0.936   &                \\		
&	159	&	0.585   &               &	162	&	0.690   &		&	165	& 0.439  & a$\rightarrow$c	\\
&	160	&	0.508   &               &	163	&	0.469 & a$\rightarrow$c	&	166	&	0.806   &b$\rightarrow$c \\		
&	161	&	0.315   &b$\rightarrow$c&	164	&	0.364   &b$\rightarrow$c&	167	&	0.806   &	\\	
&	162	&	0.471   &a$\rightarrow$c&	169	&	0.403   &b$\rightarrow$c&	168	&	0.800   &	\\		
&	170	&	0.343	&b$\rightarrow$c&	170	&	0.365   &               &	169	&	0.714   &       \\		
&	172	&	0.480   &b$\rightarrow$c&    	& $\mathbf{Z=123}$      &	    	&	170	&	0.551	&        \\		
&	173	&	0.501   &               &	159	&	0.831   &               &		&$\mathbf{Z=126}$&	\\		
&	174	&	0.400   &               &	160	&	0.821   &               &	162	&	0.995   &                \\		
&		& $\mathbf{Z=120}$	&       &	161	&	0.863	&               &	163	&	1.099   &	\\		
&	156	&	0.613   &       	&	162	&	0.924   &       	&	164	&	1.034   &        	\\		
&	157	&	0.731   &       	&	163	&	0.496	&a$\rightarrow$c&	165	&	0.802   &		\\		
&	158	&	0.652   &       	&       164	&       0.480	&a$\rightarrow$c&	166	&	0.912   &		\\		
&	159	&	0.778	&       	&	168	&	0.357   &b$\rightarrow$c&	167	&	0.807	&	        \\		
&	160	&	0.696	&         	&	169	&	0.300   &b$\rightarrow$c&	168	&	0.845	&		\\		
&	161	&	0.658   &	        &	&$\mathbf{Z=124}$	&		&       169	&       0.911	&		\\
&	162	&	0.581   &a$\rightarrow$c&	160	&	0.819	&       	&	170	&	0.735   &		\\		
&	163	&	0.323   &a$\rightarrow$b&	161	&	0.868   &		&	171	&	0.534	&		\\		
&		&$\mathbf{Z=121}$&              &	162	&	0.896   &        	&	172	&	0.434   &		\\		
&	157	&	0.747	&	    	&	163	&	0.741   &        	&	   	&	    	&	    	\\		
&	158	&	0.774   &         	&	164	&	0.739   &a$\rightarrow$c&	        &	    	&	    	\\		
&	159	&	0.690   &	    	&	165	&	0.333	&b$\rightarrow$c&         	&	    	&	    	\\		
&	160	&	0.830   &	    	&	166	&	0.334   &b$\rightarrow$c&	   	&	    	&	    	\\		
&	161	&	0.688  	&        	&	167	&	0.455   &b$\rightarrow$c&	        &        	&	    	\\		
&	162	&	0.633   &b$\rightarrow$c&	168	&	0.519   &b$\rightarrow$c&               &	    	&	     \\		
&	   	&	    	&	    	&	169	&	0.459	&         	&	   	&	    	&	    	\\		
&       	&	        &	    	&	170	&	0.328	&	    	&	   	&	    	&	    	\\		
\noalign{\smallskip}

\end{tabular}

\end{ruledtabular}

\end{table*}


\endgroup

 \subsection{Isotopic dependence}

\begin{figure}
 \includegraphics[width=1.2\linewidth,height=3.2in]{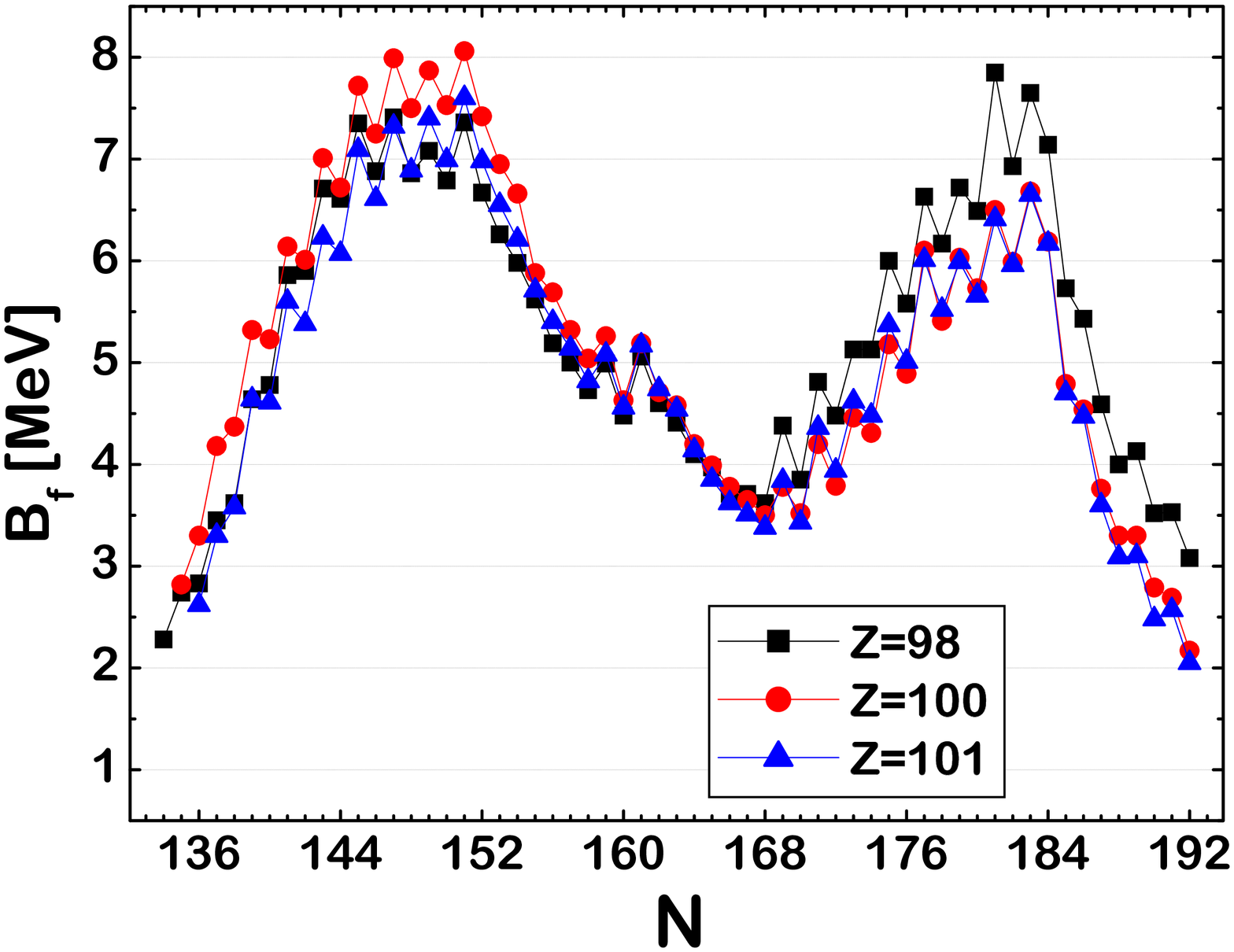}
 \caption{Isotopic dependance of fission barriers for $Z=98,99$ and $Z=100$.}
  \label{figb98-100}
 \end{figure}

\begin{figure}
 \includegraphics[width=1.2\linewidth,height=3.2in]{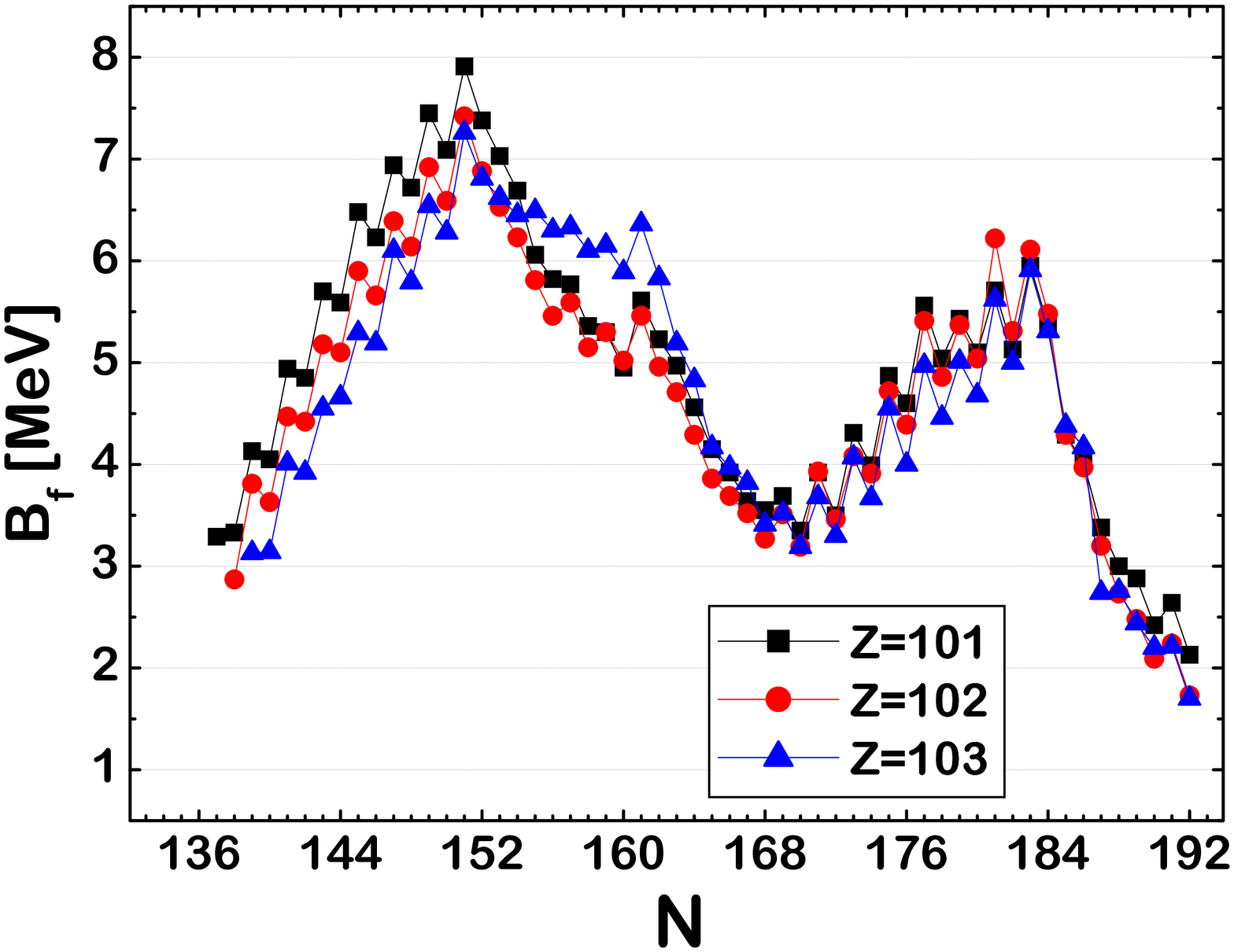}
 \caption{The same as in Fig.\ref{figb98-100} but for $Z=101,102$ and $Z=103$.}
  \label{figb101-103}
 \end{figure}

\begin{figure}
 \includegraphics[width=1.2\linewidth,height=3.2in]{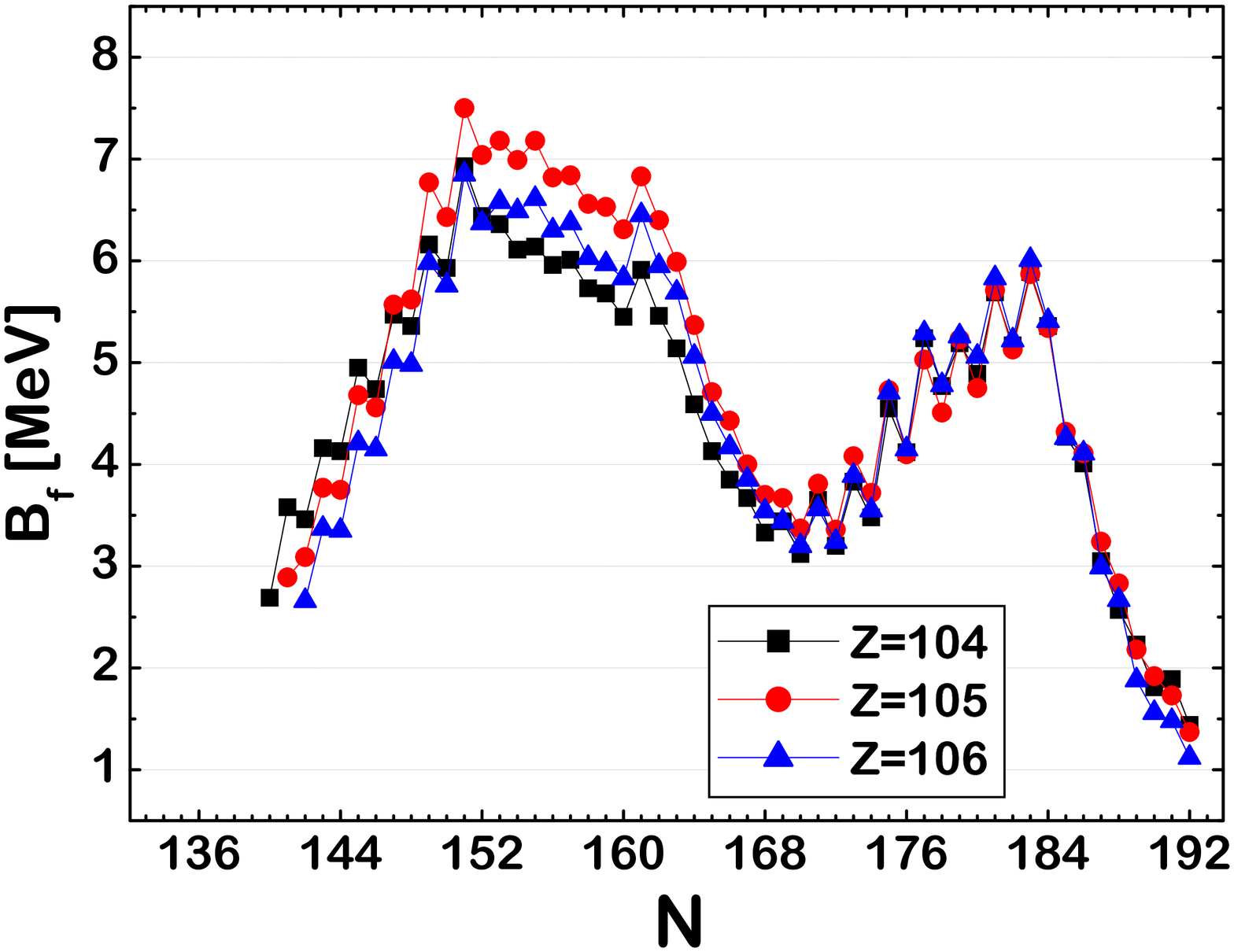}
 \caption{The same as in Fig.\ref{figb98-100} but for $Z=104,105$ and $Z=106$.}
  \label{figb104-106}
 \end{figure}

\begin{figure}
 \includegraphics[width=1.2\linewidth,height=3.2in]{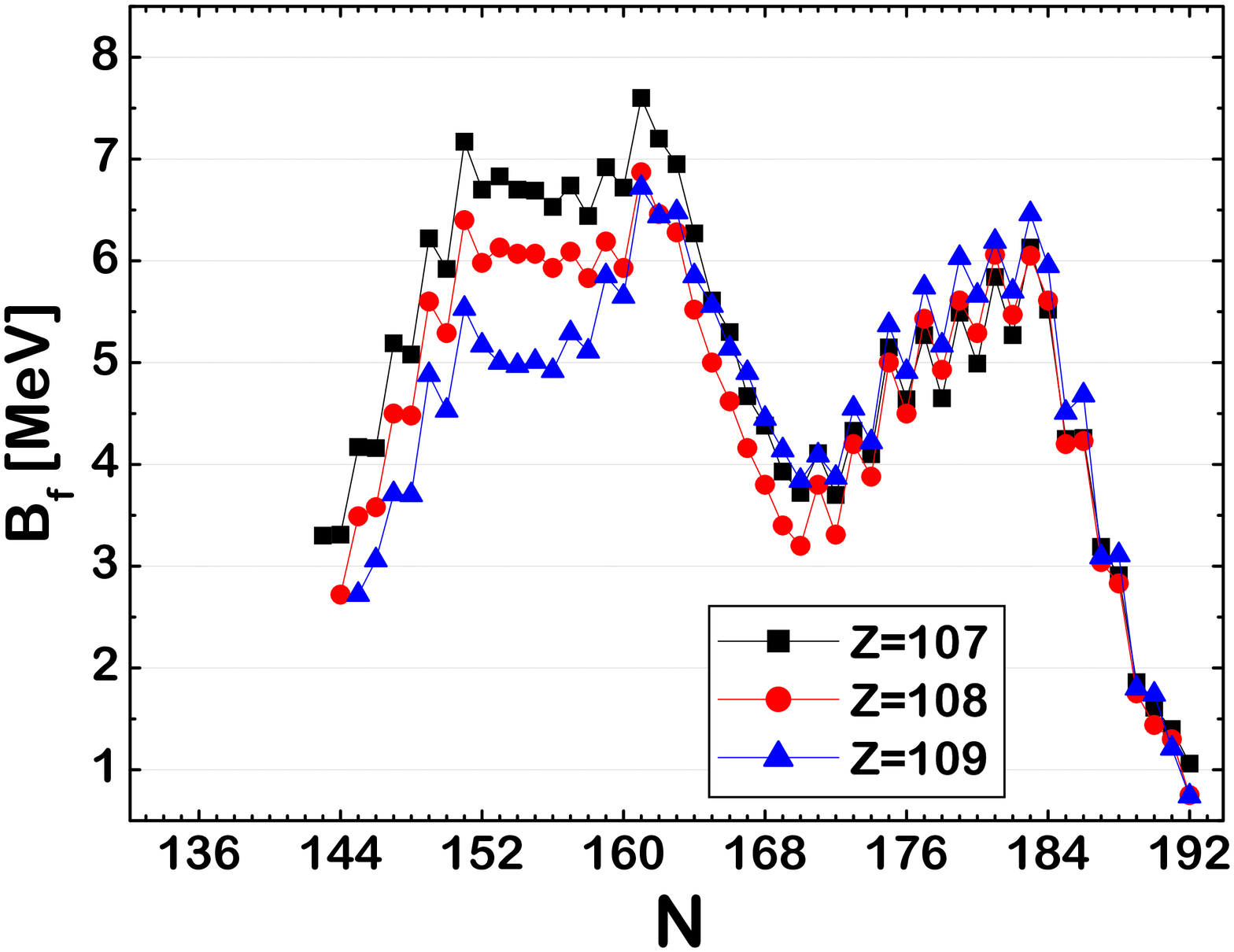}
\caption{The same as in Fig.\ref{figb98-100} but for $Z=107,108$ and $Z=109$.}
  \label{figb107-109}
\end{figure}

\begin{figure}
 \includegraphics[width=1.2\linewidth,height=3.2in]{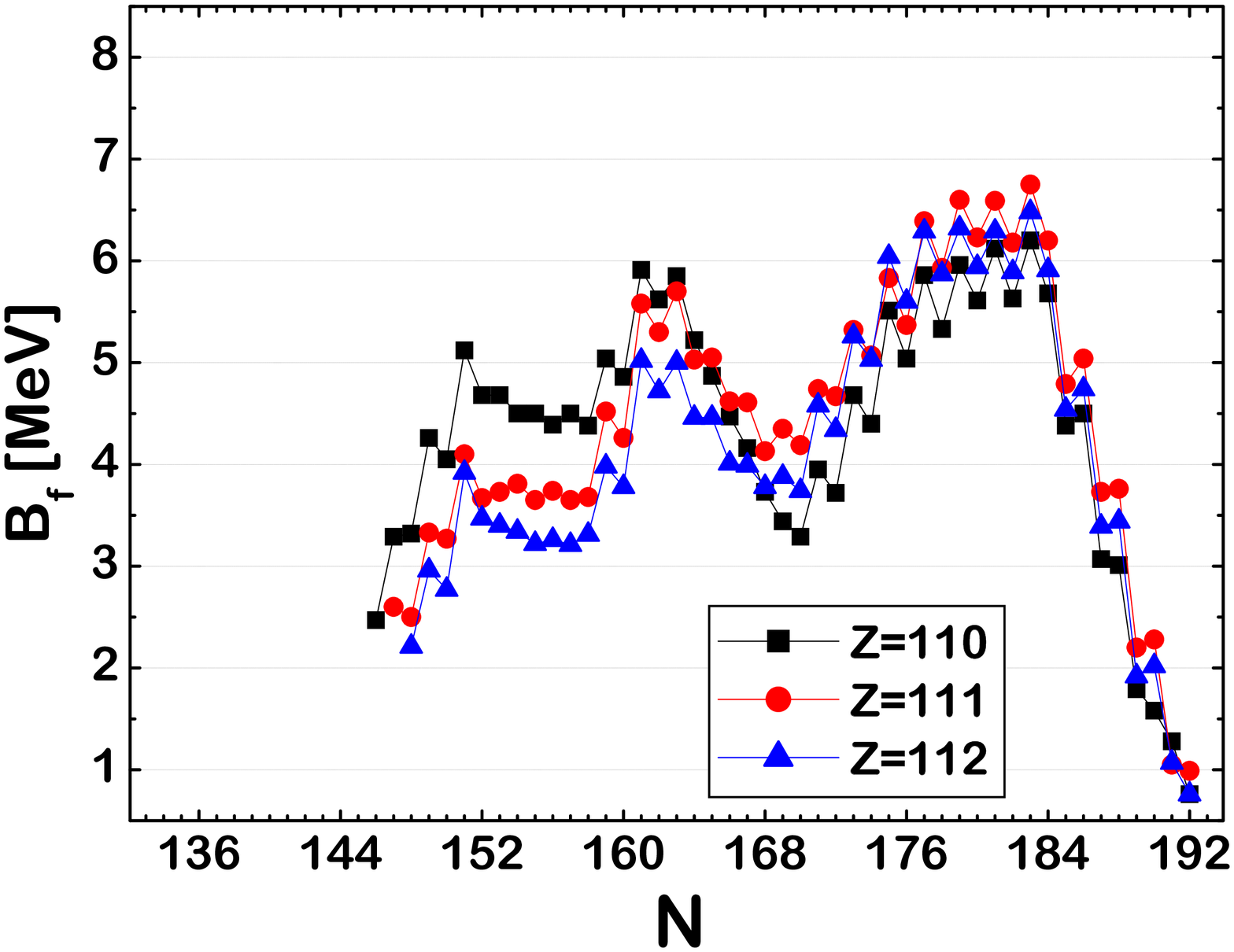}
\caption{The same as in Fig.\ref{figb98-100} but for $Z=110,111$ and $Z=112$.}
 \label{figb110-112}
\end{figure}

\begin{figure}
 \includegraphics[width=1.2\linewidth,height=3.2in]{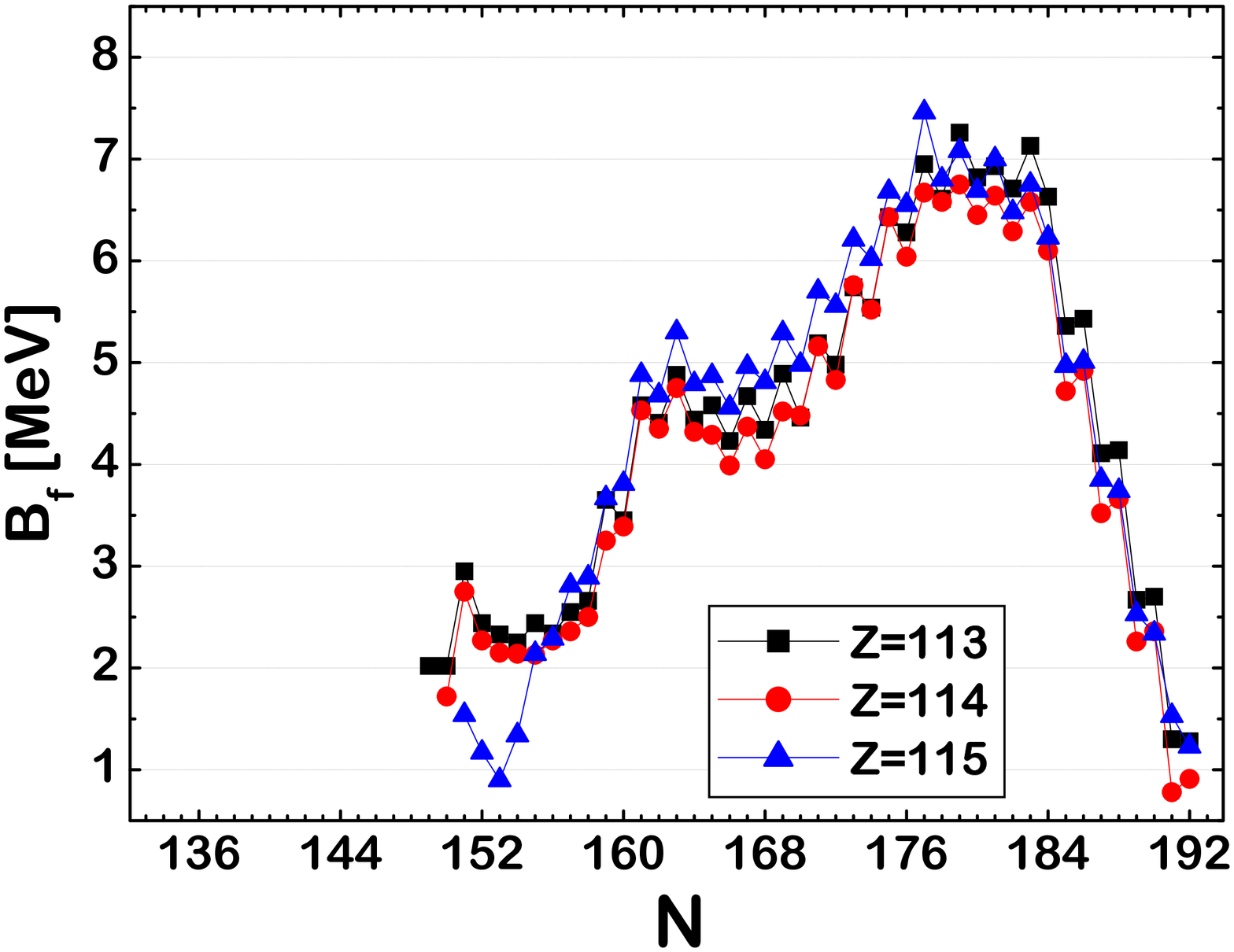}
\caption{The same as in Fig.\ref{figb98-100} but for $Z=113,1143$ and $Z=115$.}
 \label{figb113-115}
\end{figure}

\begin{figure}
 \includegraphics[width=1.2\linewidth,height=3.2in]{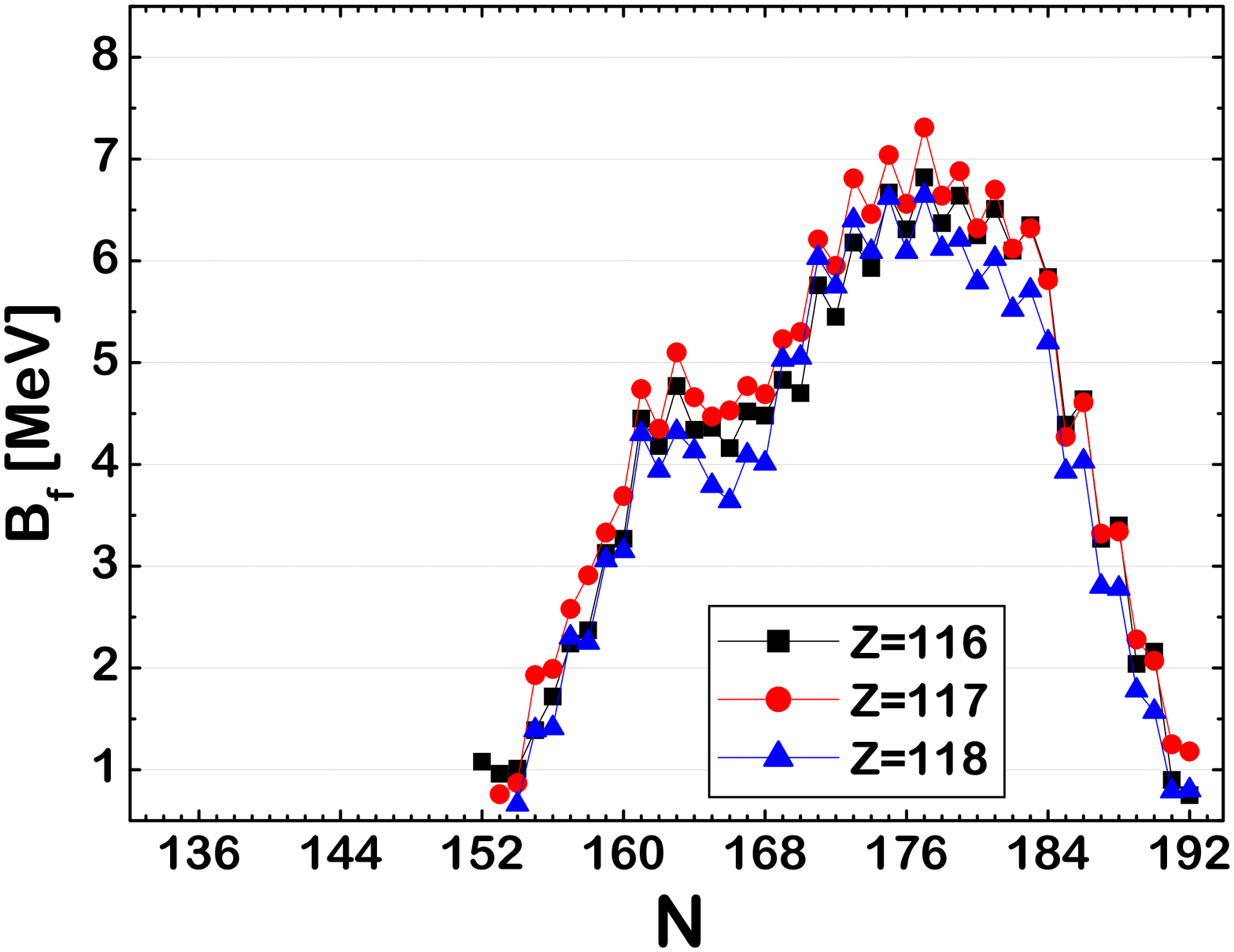}
\caption{The same as in Fig.\ref{figb98-100} but for $Z=116,117$ and $Z=118$.}
 \label{figb116-118}
\end{figure}

\begin{figure}
 \includegraphics[width=1.2\linewidth,height=3.2in]{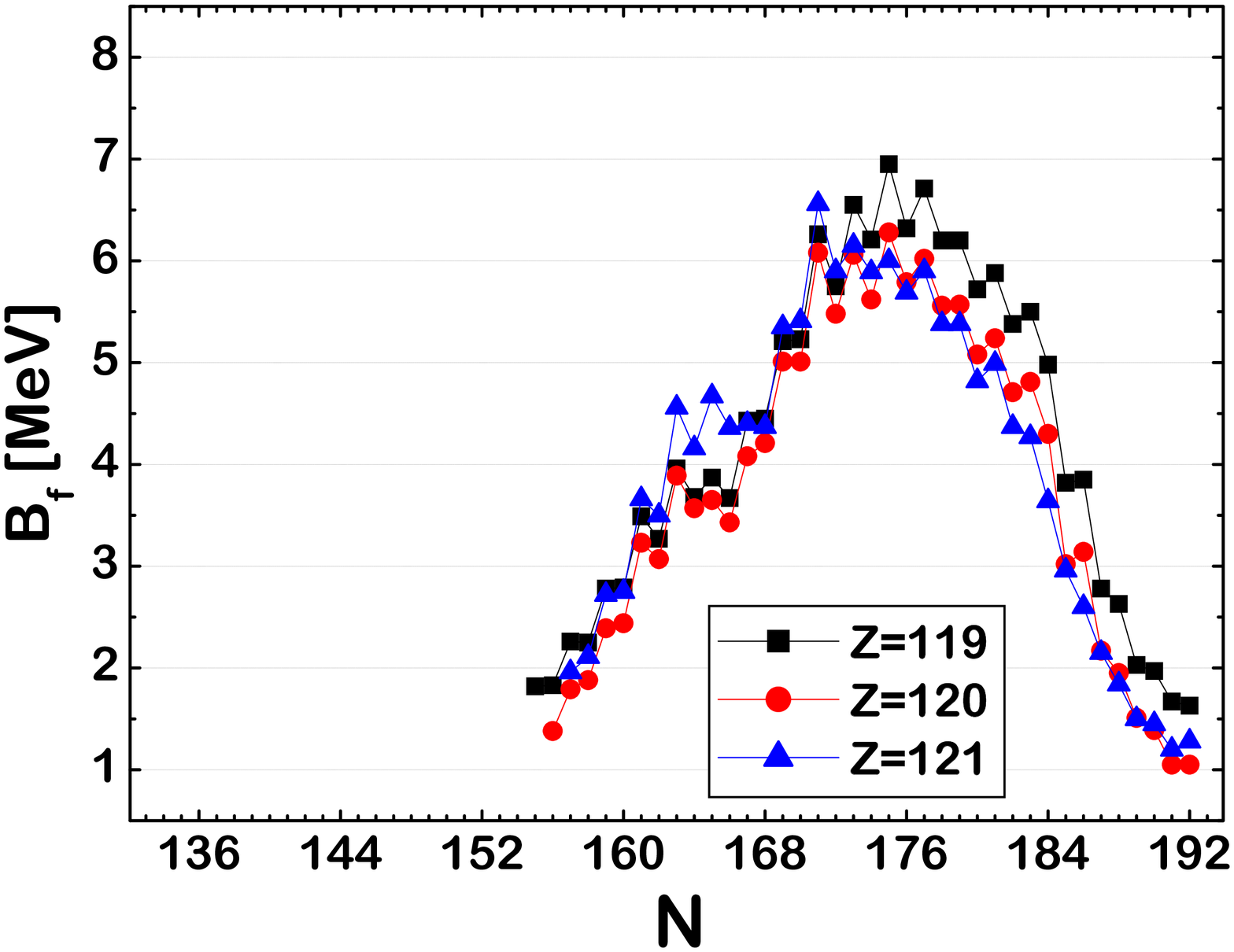}
\caption{The same as in Fig.\ref{figb98-100} but for $Z=119,120$ and $Z=121$.}
 \label{figb119-121}
\end{figure}

\begin{figure}
 \includegraphics[width=1.2\linewidth,height=3.2in]{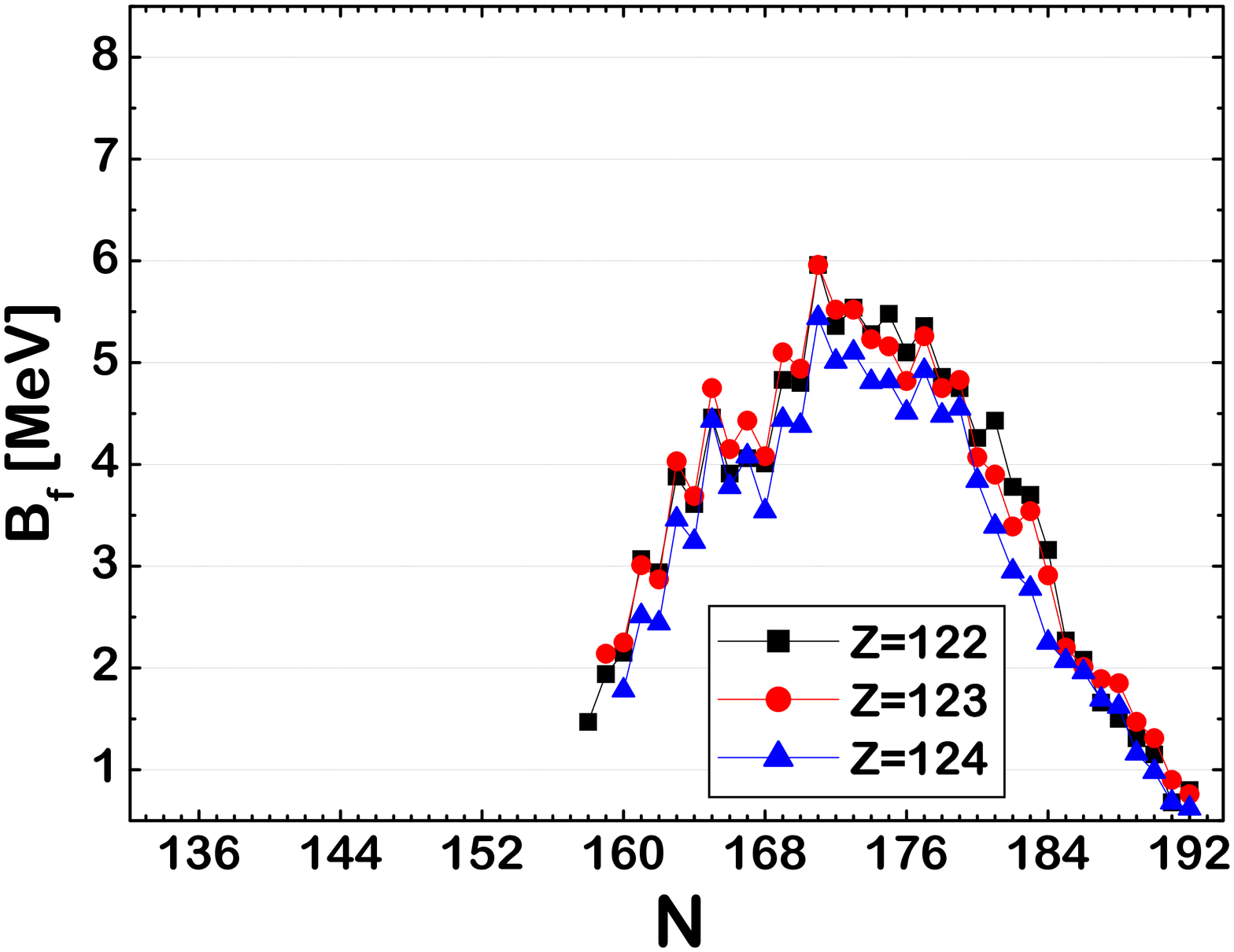}
\caption{The same as in Fig.\ref{figb98-100} but for $Z=122,123$ and $Z=124$.}
 \label{figb122-124}
\end{figure}

\begin{figure}
\includegraphics[width=1.2\linewidth,height=3.2in]{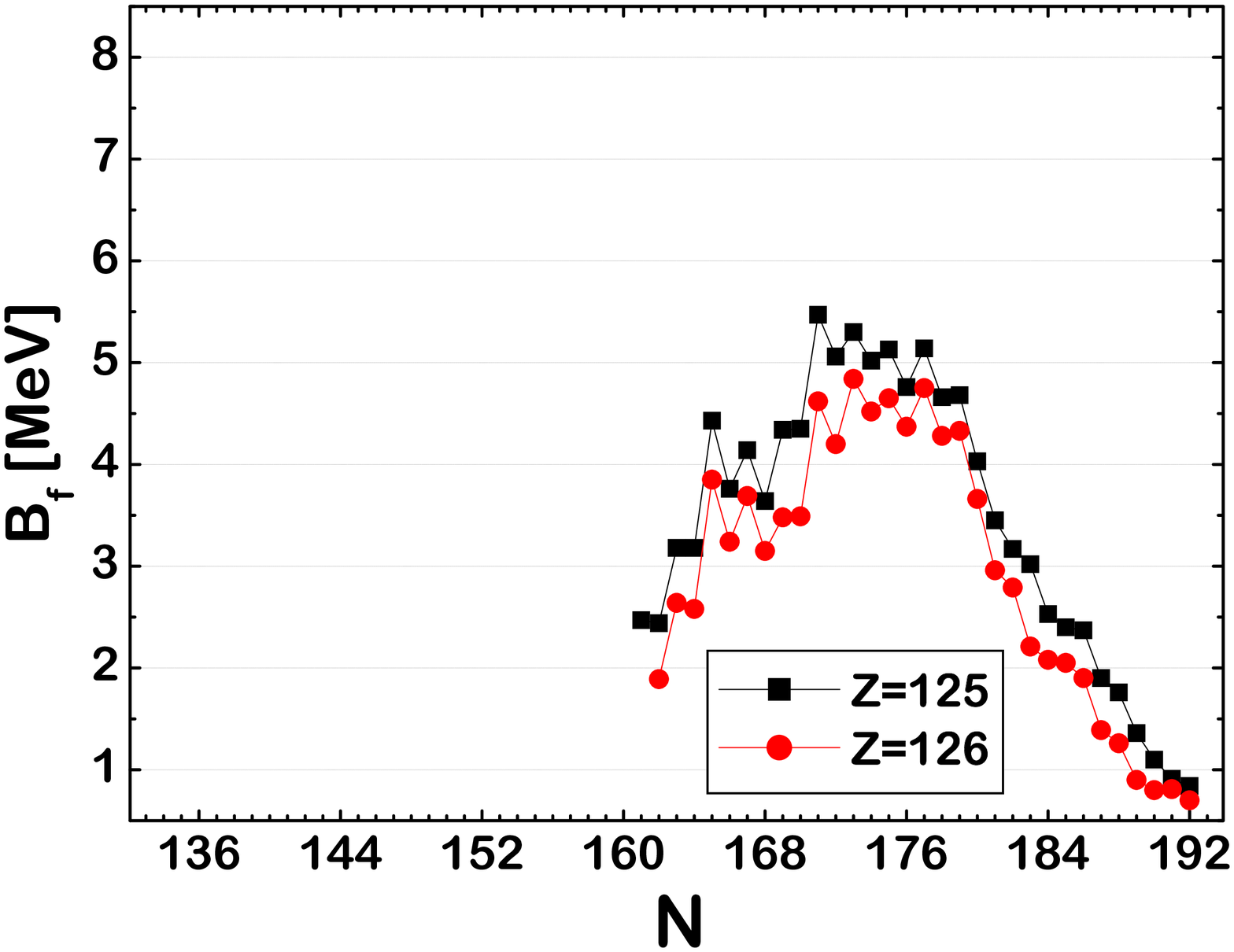}
 \caption{The same as in Fig.\ref{figb98-100} but for $Z=125$ and $Z=126$.}
  \label{figb125-126}
\end{figure}

 Calculated fission barriers given in Table \ref{bartot} are illustrated along
 isotopic chains in Figures: \ref{figb98-100} - \ref{figb125-126}.
 Generally, it can be seen that:
i) in the whole region $Z=$98 - 126 the fission barrier heights are limited
 by: $B_f\leq 8.06$ MeV;
ii) there are characteristic maxima of fission barriers at $Z\approx 100$,
 $N\approx 150$, near $Z=108$, $N=162$
 (deformed magic shells) and $Z=114$, $N=178$ (not 184); high barriers
 occur also at the border of the studied region, for $Z=98$, $N\approx 183$;
iii) over intervals of $N$ where $B_f(N)$ increase or are on average constant,
 the fission barriers in a neighboring system $N_{even}+1$ are higher
 than $B_f(N_{even})$; it may the opposite over intervals where $B_f(N)$
 strongly decrease; the same behaviour can be seen when comparing barriers for
 isotones - see Fig. \ref{Gp169}. This quite pronounced odd-even staggering in
 barriers is related to a decrease in the pairing gap due to blocking as it
 will be discussed in the next subsection.

 In the isotopic dependence of the fission barriers for Cf, Es and Fm nuclei,
  shown in Fig. \ref{figb98-100}, there are two
 peaks of a similar size, at $N =$ 152 and
 $N =$ 184. The minima of $B_f(N)$ occur at $N\approx 170$. Odd-even staggering
 in $B_f$ for Es is stronger around $N=152$, while for Cf
 it is stronger near $N=184$.


For Md, No and Lr isotopes (Fig. \ref{figb101-103}), the second
 maximum around $N=$184 is weakening. A maximum associated with the semi-magic
 deformed shell at $N=$162 appears. As before, the minima of $B_f(N)$ are
 located at $N\approx 170$.
 For Rf, Db, Sg, Bh and Hs nuclei (Fig. \ref{figb104-106} and
 \ref{figb107-109}), previously distinct maximum
 at $N=$152 becomes more flat, and a kind of plateau forms
  between $N=$152 and 162. For Mt isotopes
 this plateau changes into a local minimum in the isotopic dependence $B_f(N)$,
 located around $N=$155.
 The highest barriers in Bh, Hs and Mt isotopic sets occur at $N\approx 162$.

For Ds, Rg and Cn nuclei (Fig. \ref{figb110-112}), with the
 increasing proton number the $N=$184 spherical shell starts to dominate.
 However, not much lower barriers are obtained near the deformed gap $N=$162.

For nuclei: $Z=$ Nh, Fl, Mc (Fig. \ref{figb113-115}), one can see
  one region with high barriers, around $N=$180. One can notice that the
 maxima in $B_f(N)$  are already shifted toward $N < 184$. Slight residues
 of the formerly observed shells at $N=$152 and $N=$162 can be spotted.

For nuclei: $Z=$ Lv, Ts, Og (Fig. \ref{figb116-118}),  the main
 maximum in $B_f$ progresses further towards smaller $N$, reaching finally
 $N\approx 175$. The minima in $B_f(N)$, observed before at $N=172$,
 gradually disappear.
For nuclei: $Z=$119, Z=120, Z=121 (Fig. \ref{figb119-121}),
 the situation is similar to that described above.
 Barriers in nuclei $Z=$122, 123, 124 (Fig. \ref{figb122-124}),
 compared to the previous set, are clearly lower.
  The maximum is even more shifted towards smaller $N$.
 For nuclei: $Z=$125, 126 (Fig. \ref{figb125-126}) the fission barriers
 are still lower. Their maxima occur at $N=171$ and 173.

\begin{figure}[t!]
 \includegraphics[width=1.2\linewidth,height=5.2in]{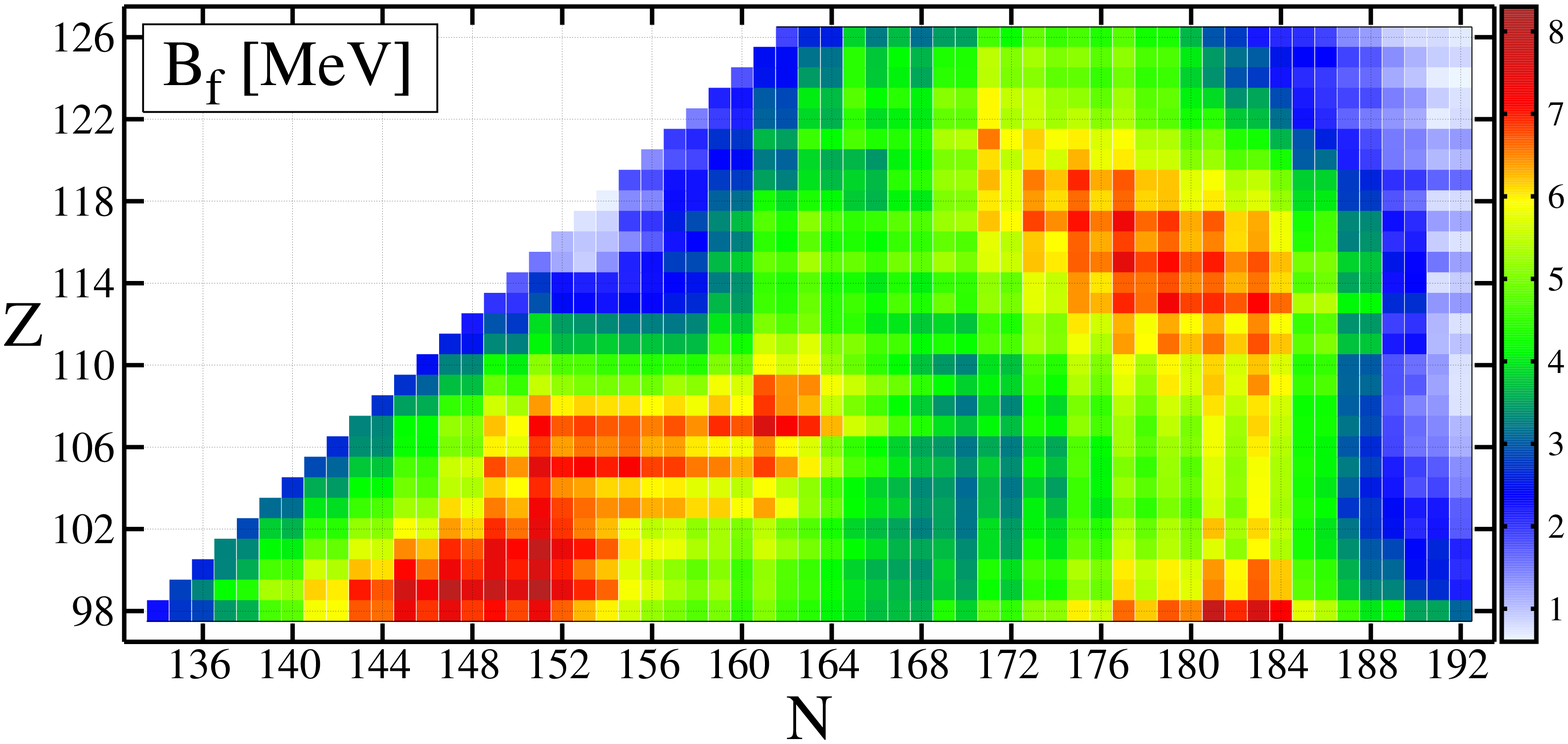}
 \caption{ Calculated fission barrier heights $B_{f}$ for superheavy nuclei.}
 \label{Bftot}
\end{figure}

All calculated fission barriers heights are collected together and shown as
 a map $B_f(Z,N)$ in Fig. \ref{Bftot}.
 One can see three areas with clearly raised barriers: around $N\approx$152,
 $N=$162 and $N\approx 180$, and the region of low barriers around $N=170$,
 as discussed above. The effect of the odd particle, i.e.
  an often (but not always) higher barrier in a neighboring odd-particle
 system can be also seen in Fig. \ref{Bftot}.
\onecolumngrid
\hspace{-10cm}
\twocolumngrid

 \subsection{Role of the pairing interaction and the odd-even barrier
  staggering}

\begin{figure}[h!]
 \includegraphics[width=1.2\linewidth,height=3.5in]{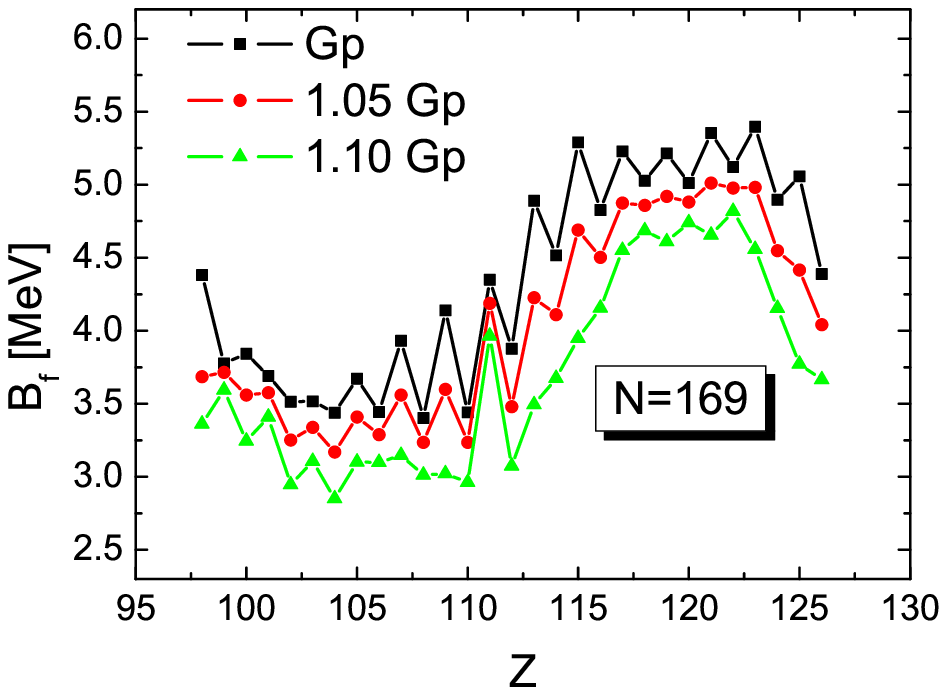}
 \caption{ Effect of the pairing strength increase
 (while keeping fixed the g.s. and saddle deformations)
 in N=169 isotones:
  standard $G_n$ and $G_p$ - black points, $G_n$ and $G_p$ increased
 by 5\% (10\%) for odd-$Z$ and odd-$N$ nuclei - red (green) points. }
 \label{Gp169}
\end{figure}

\begin{figure}[h!]
 \includegraphics[width=1.2\linewidth,height=3.5in]{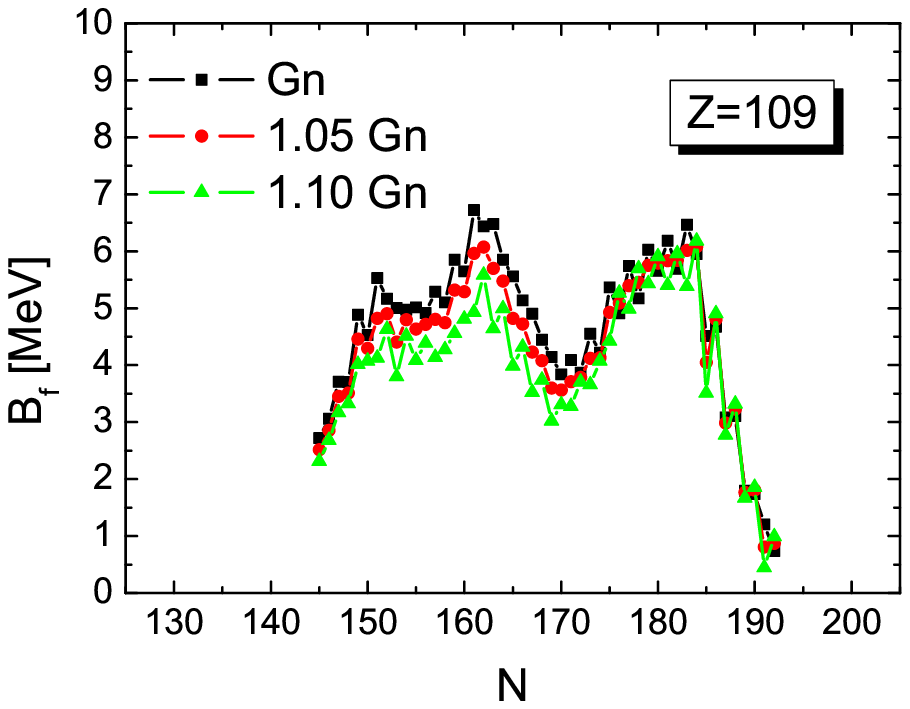}
 \caption{ Effect of the pairing strength increase
 (while keeping fixed the g.s. and saddle deformations)
 in Z=109 isotopes:
  standard $G_n$ and $G_p$ - black points, $G_n$ and $G_p$ increased
 by 5\% (10\%) for odd-$Z$ and odd-$N$ nuclei - red (green) points. }
 \label{Gn109}
\end{figure}

 It is known that the blocking procedure often causes an excessive reduction
  of the pairing gap in systems with an odd particle number. This effect
 is much more pronounced in the g.s. than in the fission saddle, as the
 pairing gap is never small in the latter. One device to avoid
  an excessive even-odd staggering in nuclear binding was to assume
  a stronger (typically by $\sim$ 5\%) pairing interaction for
  odd-particle-number systems, see \cite{Gor0,Gor1,Gor2,Gor3}.
  Here, instead of performing another grid calculation with modified pairing
 strengths, we tested the magnitude of their effect on fission barriers by
  increasing them by 5 and 10 percent for odd particle numbers (neutrons
  or protons) at previously found ground states and saddle points.
  The results of this test are presented in Fig. \ref{Gp169} for  the N=169
 isotones and in Fig. \ref{Gn109} for the Z=109 isotopic chain.

 Both the isotopic and isotonic dependence show that increasing
 the intensity of pairing leads to a reduction of the fission barrier by a
 variable amount.
   When the pairing strengths are increased by 5\% for odd particle numbers,
  the fission barriers decrease in odd-even, even-odd and odd-odd systems
  by up to 0.5 MeV; the 10\% increase in the pairing strengths
   can decrease the barriers at most by about 1 MeV.
  The same pairing change leads to the suppression, and then the inversion
  of the staggering effect.

 The even-odd barrier staggering related to pairing is convoluted with the
 isotopic or isotonic dependence related to the mean-field.
   With the original pairing, when one separates a linear part of the latter
   by calculating: $B_f(Z_{odd},N)-1/2[B_f(Z_{odd}+1,N)+
  B_f(Z_{odd}-1,N)]$, and an analogous quantity for odd neutron numbers,
  one obtains numbers between 1.053 and $-0.947$ MeV, with the average
  of $\approx 0.22$ MeV for protons and $\approx 0.26$ MeV for neutrons.
  As shown by black points in Fig. \ref{Gp169}, \ref{Gn109}, the effect is
  indeed irregular and, when present, typically at the level of several
  hundred keV.


  The 5\% increase in pairing for odd particle numbers
 reduces the staggering in $N=169$ isotones and nearly cancels it in
  $Z=109$ isotopes (red points in Fig.\ref{Gp169} and \ref{Gn109}).
  The important point is that the 10\% increase in pairing for odd number of
  particles {\it inverts} the staggering, at least locally: near $Z=120$ in
  $N=169$ isotones and near $N=153$, $N=162$ and $N=180$ in Mt isotopes
  (green points in Fig.\ref{Gp169} and \ref{Gn109}).

  Although the spontaneous fission rates of odd-particle number
  nuclei are smaller by 3-5 orders of magnitude than those of their even
  neighbors, the experimental fission barriers in actinides show only
  a moderate odd-even staggering, c.f. \cite{SMIR93,OBN}. Still, it is
 inconceivable that the fission barriers in odd-$Z$ or odd-$N$ systems should
 be on average {\it smaller} than in their even neighbors.
  This indicates that the 10\% increase in pairing strengths in odd-$N$
  or odd-$Z$ systems would be too large.
  A qualitative argument which follows is that even if the blocking method
 overestimates the pairing decrease, the fission barriers of odd-$Z$ or/and
  odd-$N$ nuclei should fall in a strip between the black and red points in
  Fig.\ref{Gp169} and \ref{Gn109}. Thus, the test of
  the pairing influence on barriers points that a possible overestimate of
   barriers in odd-$A$ and odd-odd nuclei, induced by the blocking, should not
   be much larger than 0.5 MeV. One may add in this context that
  the barriers from the FRLDM model do not show any even-odd staggering
  due to the way the pairing was included there.

 \subsection{Comparison with other theoretical calculations and some empirical
    data}

 Let us discuss the results in Table \ref{bartot} in relation to
 available empirical data and to the other theoretical estimates.

  As an empirical check of our model, one can use the barriers
 in the actinide region. We have reported quite a spectacular agreement
   of the calculated first \cite{Kow} and second \cite{IIbarriers} fission
 barriers in even-even actinides with the data \cite{SMIR93,OBN}, with root mean square
 deviation 0.5 MeV and 0.7 MeV, respectively.

The heaviest nucleus in which the fission barrier height has been measured
 recently is $^{254}$No.
 The value $B_f$=6.0 $\pm$ 0.5 MeV at spin 15$\hbar$, giving by
 extrapolation $B_f$=6.6 $\pm$ 0.9 MeV at the spin 0$\hbar$, has been
deduced from the measured distribution of entry points in the excitation
 energy vs. angular momentum plane \cite{seweryniak}. This result perfectly
 agrees with our evaluation: $B_f$=6.88 MeV (at spin 0$\hbar$ ) and with the
 MM model \cite{FRLDM} which gives: $B_f$=6.76 MeV.
 The selfconsistent calculations, mainly based on the Skyrme interaction,
 overestimate this barrier significantly \cite{Bonneau,Duguet,staszczak2006}
 (9.6 and 8.6 or 12.5 MeV, respectively).
 There are experimental estimates of barriers in a few SH nuclei,
 based on observed ER production probabilities \cite{ITKIS}, which again
 well agree with our barriers, see  \cite{Kow}. Apart from those,
 fission barriers in the SH region are generally unknown.

 As a supplementary insight, one can crosscheck barriers evaluated
 within various models. Quite recently we noted a dramatic divergence in
 calculated fission barriers \cite{japan2015}.
 Since, as it was discussed previously, the inclusion of traxiality is
 absolutely necessary in the SH region, we have
 chosen only models which take this into account.
In fact, there is only one systematic calculation, including triaxiality and
   odd-particle-number nuclei - the Finite Range Liquid Drop Model
  \cite{mol95,Moller2009,FRLDM} (FRLDM) developed by Los Alamos group.
 It can be noted though, that the inner fission barrier is fixed there
 in only three-dimensional deformation space, what is certainly not enough.

 The first conclusion from the comparison between our results and those of
 FRLDM is that a conspicuous barrier staggering between odd- and even-particle
 number nuclei is obtained in the Woods-Saxon model. As mentioned before, this
 results from the blocking treatment of pairing. At present it is not certain
  how large this staggering should be.

  One can include more models for comparison if one confines it to
 even - even nuclei. We take the covariant density functional model \cite{RMF}
  with the nonlinear meson-nucleon coupling, represented by the NL3*
 parametrization of the relativistic mean-field (RMF) Lagrangian and
 the Hartree-Fock-Bogoliubov (HFB) approach with the SkM* Skyrme energy
 density functional \cite{SKM}.

\begin{figure}[h!]
 \includegraphics[width=1.2\linewidth,height=3.5in]{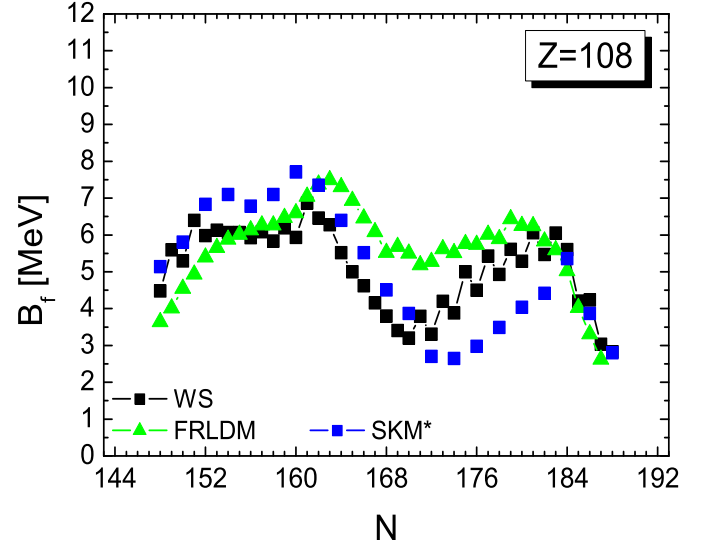}
 \caption{ Fission barriers predicted by various models for Hassium isotopes: black - WS model, green – FRLDM \cite{FRLDM}, blue – SkM* \cite{SKM}, red – RMF with NL3 parametrization \cite{RMF}. Experimental data taken from \cite{ITKIS}. (For interpretation of the references to color in this figure legend, the reader is referred to the web version of this article.)}
 \label{108th}
\end{figure}

 As can be seen in Fig. \ref{108th}, fission barriers in Hassium nuclei are
 quite similar in all models. The values of $B_f$
 differ up to 2 MeV, but never more. Regarded as a function of $N$, they show
 a maximum close to the semi- magic number $N=$162 while the second
 maximum is related with the $N=$184 spherical gap. In the FRLDM this
 maximum is barely outlined and slightly shifted to the neutron
  deficient side. The minimum in barriers is obtained in both MM
 models at the similar place ($N=$170), while the RMF gives the smallest
 barriers at $Z=$174.

\begin{figure}[h!]
 \includegraphics[width=1.2\linewidth,height=3.5in]{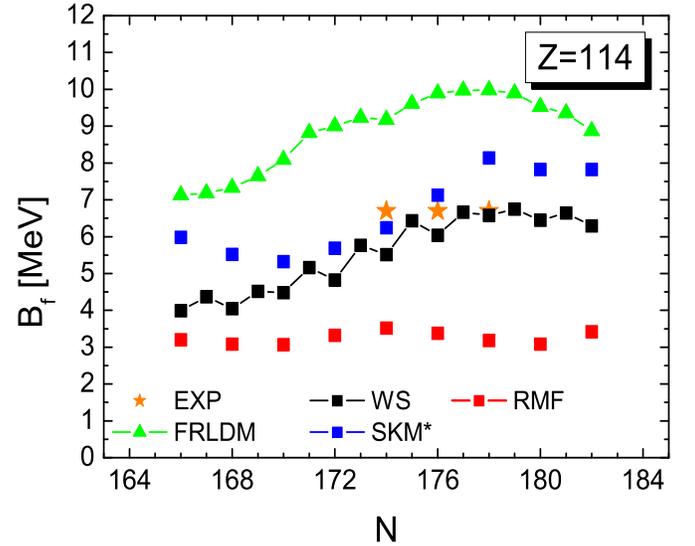}
 \caption{ The same as in Fig \ref{108th} but for Z=114.}
 \label{114th}
\end{figure}

As one can see in Fig. \ref{114th}, for Flerovium isotopes  the barriers
 calculated here are in agreement with the experimental (empirical) estimates
  \cite{ITKIS} and with the self-consistent calculations \cite{SKM} based on
 the SKM* interaction. The FRLDM \cite{FRLDM} overestimates these
 quasi-empirical barriers \cite{ITKIS} significantly.
Although only the lower limit for the barrier height has been estimated in
 \cite{ITKIS}, which would reproduce the known cross sections on the picobarn
 level, such a high barrier seems problematic, see discussion in
 \cite{Wilczynska1,Wilczynska2}.
 On the other hand, with extremely small barriers obtained within the RMF model
 one cannot explain experimentally known millisecond fission half-life
 in $^{284}$Fl.
 One should note, however, that a slight tuning of the RMF model \cite{RMF2015}
 gives higher barriers, thus, closer to ours. This is true, especially in Cn
 and Fl isotopes, see details in Fig. 5 in \cite{RMF2015} and discussion
 included there.


\begin{figure}[h!]
 \includegraphics[width=1.2\linewidth,height=3.5in]{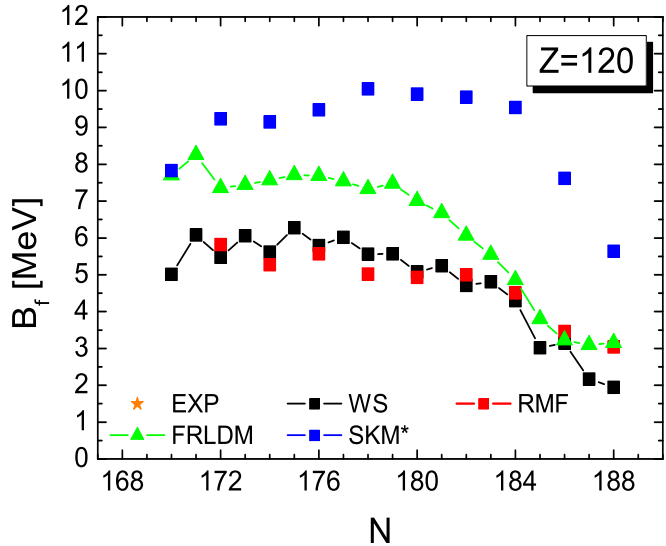}
 \caption{ The same as in Fig \ref{108th} but for Z=120.}
 \label{120th}
\end{figure}

For $Z=$120 our results, shown in Fig. \ref{120th}, are very close to those
 obtained within the RMF model.
 The results of \cite{FRLDM} are systematically higher by $\approx$ 1 MeV.
This is in an evident contrast with the Skyrme SkM* prediction \cite{SKM} of
 the highest barriers for $Z=120$ \cite{SKM} - related to the proton magic gap.
 Three models: FRLDM, RMF and ours converge at N=182-184 to
 $B_{f} \simeq 5$ MeV. The nucleus $^{302}$120 is particulary interesting,
 as two unsuccessful attempts to produce it have already taken place in GSI,
 providing a cross-section limit of 560 fb \cite{120GSI1} or 90 fb in
 \cite{120GSI2}, and in Dubna \cite{120DUBNA}, providing the limit of 400 fb.
 The cross-section estimates \cite{Wilczynska3} do not support a possibility
 of an easy production of this SH isotope in the laboratory.
It seems that with the barrier of the order of 10 MeV, as obtained in the
 frame of the self-consistent theory, producing superheavy Z=120 nuclei
  should not pose any difficulties.

\begin{figure}[h!]
 \includegraphics[width=1.2\linewidth,height=3.5in]{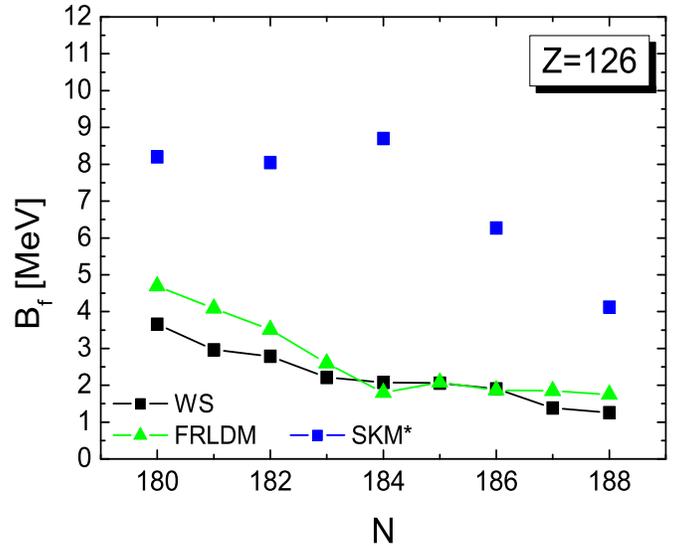}
 \caption{ The same as in Fig \ref{108th} but for Z=126.}
 \label{126th}
\end{figure}

 In the case of $Z=$126, shown in Fig. \ref{126th}, both MM models give
 significantly smaller barriers than the model based on the SKM* force.
 For example, the barrier $B_{f}\approx 9$ MeV for $^{310}126$, calculated with
 this Skyrme interaction, is still impressively large. This might induce
 thoughts on the ways of synthesis of such superheavy systems, but one has to
 remember that the predicted half-lives with respect to $\alpha$ decay
 are below the present-day $10^{-5}$ s time-limit for the experimental
 identification.
 On the contrary, $B_f\approx 2$ MeV obtained in the MM
 approach does not induce any hopes; it only points to a quite striking
 disagreement between models.

 \section{Conclusions}

 We have determined fission barriers for 1305 heavy and superheavy nuclei,
 including odd-$A$ and odd-odd systems, within the macroscopic-microscopic
 method by following the adiabatic configuration in each nucleus.
 The applied Woods-Saxon model was widely used for heavy nuclei
 and well reproduces experimental fission barriers in actinides.
 For odd-$Z$ or/and odd-$N$ nuclei pairing was included within the blocking
 procedure.
  Triaxial and mass-asymmetric deformations were included and
 the IWF method used for finding the saddles which allowed to
  escape errors inherent in the constrained minimization approach.
 To find saddles, energy for each nucleus was calculated on a 5D deformation
  grid and then 5-fold interpolated in each dimension for
  the IWF search. Two additional energy grids: a second 5D and another 7D,
   were calculated in order to include nonaxial hexadecapole and
  mass-asymmetry effects on fission barriers.
 The following conclusions can be drawn from our investigation:

i)    Global calculations confirm the existence of two physically important
   areas in the $Z$-$N$ plane with prominent barriers: one located around
  the semi-magic quantum numbers $Z=100-108$ and $N=150-162$ (connected with
   deformed closed shells) and the second - of nearly spherical nuclei around
 $Z=114$ and $N=176-180$.
   The highest fission barrier among the studied nuclei occurs in very exotic
    Es$^{250}$.

ii)   The well-known effect of the mass asymmetry on the second barrier
      in actinides is not very relevant for the heaviest nuclei since very
  deformed barriers at $\beta_{20}\approx 0.8$ decrease with increasing $Z$
    and fission barriers are fixed by the less deformed saddles.
     However, in some nuclei with $Z\geq 109$ the mass(reflection) asymmetry
    effect lowers the first saddles which are sometimes split into two humps.
   It seems that this concerns only axially-symmetric saddles. The largest
    barrier lowering (by 0.8 MeV) has been observed for $Z=113$ and $N=157$.

iii)  It has been demonstrated that the inclusion of triaxial shapes
      significantly reduces the fission barriers by up to 2.5 MeV;
      about 70\% of the found fission barriers correspond to triaxial
      saddles.
      Besides the quadrupole nonaxiality we checked also the effect of
     hexadecapole nonaxiality which significantly lowers the fission
     barrier in $Z\geq 119$ nuclei, especially neutron-deficient ones.

iv)   Rather strong, irregular odd-even $Z$ or $N$ barrier staggering effect
    resulted from the blocking formalism used for pairing.
      The barrier of an odd nucleus $Z_{even}+1$ or $N_{even}+1$ is typically
   by several hundred keV higher than that of its even neighbor.

v)    The existing theoretical evaluations of fission barriers differ
     significantly. Even the results of the two models based on the
   microscopic-macroscopic approach differ dramatically for some nuclei.
   Our calculations indicate, in contrast to the self-consistent mean-field
  studies, that fission barriers, still quite substantial for
    some $Z=118$ nuclei, become lower than 5.5 MeV for $Z=126$.


\section{ACKNOWLEDGMENT}
 M.K. and J.S. were co-financed by the National Science Centre under Contract No. UMO-2013/08/M/ST2/00257  (LEA COPIGAL). One of the authors
(P.J.) was cofinanced by Ministry of Science and Higher Education:
„Iuventus Plus” grant Nr IP2014 016073.
This research was also supported by an allocation
of advanced computing resources provided by the Swierk Computing Centre (CIS)
 at the National Centre for Nuclear Research (NCBJ) (http://www.cis.gov.pl).

\begingroup

\squeezetable


\begin{table*}

\caption{\label{bartot} Calculated fission barrier heights (in MeV).}

 \begin{ruledtabular}

\begin{tabular}{|cccc|ccc|ccc|ccc|ccc|}

&     N &   A  &  $B_{f}$   &      N    &    A  &  $B_{f}$  &   N       &   A   &  $B_{f}$  &        N  &    A  &  $B_{f}$  &        N  &   A   &  $B_{f}$ \\

\noalign{\smallskip}\hline\noalign{\smallskip}


 &       & \textbf{Z=98}&    &           & \textbf{Z=99}&   &           &  \textbf{Z=100}&  &          &  \textbf{Z=101}&  &           & \textbf{Z=102}&   \\

	&	134	&	232	&	2.28	&		&		&		    &		&		&	    	&		&	  	&	    	&		&		&		     	\\
	&	135	&	233	&	2.74	&	135	&	234	&	2.82	&		&		&		    &		&		&	     	&		&		&		    	\\
	&	136	&	234	&	2.83	&	136	&	235	&	3.30	&	136	&	236	&	2.62	&		&		&	    	&		&		&		    	\\
	&	137	&	235	&	3.45	&	137	&	236	&	4.18	&	137	&	237	&	3.30	&	137	&	238	&	3.29	&		&		&		    	\\
	&	138	&	236	&	3.62	&	138	&	237	&	4.37	&	138	&	238	&	3.58	&	138	&	239	&	3.33	&	138	&	240	&	2.87		\\
	&	139	&	237	&	4.64	&	139	&	238	&	5.32	&	139	&	239	&	4.64	&	139	&	240	&	4.13	&	139	&	241	&	3.81		\\
	&	140	&	238	&	4.78	&	140	&	239	&	5.23	&	140	&	240	&	4.61	&	140	&	241	&	4.05	&	140	&	242	&	3.63		\\
	&	141	&	239	&	5.86	&	141	&	240	&	6.14	&	141	&	241	&	5.60	&	141	&	242	&	4.94	&	141	&	243	&	4.47		\\
	&	142	&	240	&	5.90	&	142	&	241	&	6.01	&	142	&	242	&	5.38	&	142	&	243	&	4.85	&	142	&	244	&	4.42		\\
	&	143	&	241	&	6.71	&	143	&	242	&	7.01	&	143	&	243	&	6.23	&	143	&	244	&	5.70	&	143	&	245	&	5.18		\\
	&	144	&	242	&	6.61	&	144	&	243	&	6.72	&	144	&	244	&	6.07	&	144	&	245	&	5.59	&	144	&	246	&	5.10		\\
	&	145	&	243	&	7.35	&	145	&	244	&	7.72	&	145	&	245	&	7.09	&	145	&	246	&	6.48	&	145	&	247	&	5.90		\\
	&	146	&	244	&	6.88	&	146	&	245	&	7.25	&	146	&	246	&	6.61	&	146	&	247	&	6.23	&	146	&	248	&	5.66		\\
	&	147	&	245	&	7.41	&	147	&	246	&	7.99	&	147	&	247	&	7.32	&	147	&	248	&	6.94	&	147	&	249	&	6.39		\\
	&	148	&	246	&	6.86	&	148	&	247	&	7.50	&	148	&	248	&	6.89	&	148	&	249	&	6.72	&	148	&	250	&	6.14		\\
	&	149	&	247	&	7.08	&	149	&	248	&	7.87	&	149	&	249	&	7.40	&	149	&	250	&	7.45	&	149	&	251	&	6.92		\\
	&	150	&	248	&	6.79	&	150	&	249	&	7.53	&	150	&	250	&	6.99	&	150	&	251	&	7.09	&	150	&	252	&	6.59		\\
	&	151	&	249	&	7.36	&	151	&	250	&	8.06	&	151	&	251	&	7.60	&	151	&	252	&	7.91	&	151	&	253	&	7.42		\\
	&	152	&	250	&	6.67	&	152	&	251	&	7.42	&	152	&	252	&	6.98	&	152	&	253	&	7.38	&	152	&	254	&	6.88		\\
	&	153	&	251	&	6.26	&	153	&	252	&	6.95	&	153	&	253	&	6.55	&	153	&	254	&	7.03	&	153	&	255	&	6.53		\\
	&	154	&	252	&	5.98	&	154	&	253	&	6.66	&	154	&	254	&	6.21	&	154	&	255	&	6.69	&	154	&	256	&	6.23		\\
	&	155	&	253	&	5.62	&	155	&	254	&	5.88	&	155	&	255	&	5.71	&	155	&	256	&	6.06	&	155	&	257	&	5.81		\\
	&	156	&	254	&	5.19	&	156	&	255	&	5.69	&	156	&	256	&	5.40	&	156	&	257	&	5.82	&	156	&	258	&	5.46		\\
	&	157	&	255	&	5.00	&	157	&	256	&	5.32	&	157	&	257	&	5.14	&	157	&	258	&	5.77	&	157	&	259	&	5.59		\\
	&	158	&	256	&	4.73	&	158	&	257	&	5.04	&	158	&	258	&	4.82	&	158	&	259	&	5.36	&	158	&	260	&	5.15		\\
	&	159	&	257	&	4.99	&	159	&	258	&	5.26	&	159	&	259	&	5.08	&	159	&	260	&	5.30	&	159	&	261	&	5.30		\\
	&	160	&	258	&	4.48	&	160	&	259	&	4.63	&	160	&	260	&	4.56	&	160	&	261	&	4.95	&	160	&	262	&	5.02		\\
	&	161	&	259	&	5.06	&	161	&	260	&	5.19	&	161	&	261	&	5.17	&	161	&	262	&	5.61	&	161	&	263	&	5.46		\\
	&	162	&	260	&	4.60	&	162	&	261	&	4.71	&	162	&	262	&	4.74	&	162	&	263	&	5.23	&	162	&	264	&	4.96		\\
	&	163	&	261	&	4.41	&	163	&	262	&	4.58	&	163	&	263	&	4.54	&	163	&	264	&	4.97	&	163	&	265	&	4.71		\\
	&	164	&	262	&	4.10	&	164	&	263	&	4.20	&	164	&	264	&	4.14	&	164	&	265	&	4.56	&	164	&	266	&	4.29		\\
	&	165	&	263	&	3.97	&	165	&	264	&	3.99	&	165	&	265	&	3.85	&	165	&	266	&	4.15	&	165	&	267	&	3.86		\\
	&	166	&	264	&	3.71	&	166	&	265	&	3.78	&	166	&	266	&	3.62	&	166	&	267	&	3.92	&	166	&	268	&	3.69		\\
	&	167	&	265	&	3.71	&	167	&	266	&	3.65	&	167	&	267	&	3.51	&	167	&	268	&	3.64	&	167	&	269	&	3.52		\\
	&	168	&	266	&	3.62	&	168	&	267	&	3.50	&	168	&	268	&	3.38	&	168	&	269	&	3.55	&	168	&	270	&	3.27		\\
	&	169	&	267	&	4.38	&	169	&	268	&	3.78	&	169	&	269	&	3.84	&	169	&	270	&	3.69	&	169	&	271	&	3.51		\\
	&	170	&	268	&	3.85	&	170	&	269	&	3.52	&	170	&	270	&	3.43	&	170	&	271	&	3.35	&	170	&	272	&	3.19		\\
	&	171	&	269	&	4.81	&	171	&	270	&	4.20	&	171	&	271	&	4.36	&	171	&	272	&	3.92	&	171	&	273	&	3.93		\\
	&	172	&	270	&	4.48	&	172	&	271	&	3.79	&	172	&	272	&	3.94	&	172	&	273	&	3.50	&	172	&	274	&	3.46		\\
	&	173	&	271	&	5.13	&	173	&	272	&	4.46	&	173	&	273	&	4.62	&	173	&	274	&	4.31	&	173	&	275	&	4.08		\\
	&	174	&	272	&	5.13	&	174	&	273	&	4.31	&	174	&	274	&	4.48	&	174	&	275	&	3.99	&	174	&	276	&	3.91		\\
	&	175	&	273	&	6.00	&	175	&	274	&	5.18	&	175	&	275	&	5.37	&	175	&	276	&	4.87	&	175	&	277	&	4.72		\\
	&	176	&	274	&	5.58	&	176	&	275	&	4.89	&	176	&	276	&	5.01	&	176	&	277	&	4.60	&	176	&	278	&	4.39		\\
	&	177	&	275	&	6.63	&	177	&	276	&	6.10	&	177	&	277	&	6.01	&	177	&	278	&	5.56	&	177	&	279	&	5.41		\\
	&	178	&	276	&	6.17	&	178	&	277	&	5.41	&	178	&	278	&	5.52	&	178	&	279	&	5.04	&	178	&	280	&	4.86		\\
	&	179	&	277	&	6.72	&	179	&	278	&	6.03	&	179	&	279	&	5.99	&	179	&	280	&	5.43	&	179	&	281	&	5.37		\\
	&	180	&	278	&	6.49	&	180	&	279	&	5.73	&	180	&	280	&	5.66	&	180	&	281	&	5.10	&	180	&	282	&	5.04		\\
	&	181	&	279	&	7.85	&	181	&	280	&	6.50	&	181	&	281	&	6.41	&	181	&	282	&	5.71	&	181	&	283	&	6.22		\\
	&	182	&	280	&	6.93	&	182	&	281	&	5.99	&	182	&	282	&	5.96	&	182	&	283	&	5.13	&	182	&	284	&	5.31		\\
	&	183	&	281	&	7.65	&	183	&	282	&	6.68	&	183	&	283	&	6.65	&	183	&	284	&	5.95	&	183	&	285	&	6.11		\\
	&	184	&	282	&	7.14	&	184	&	283	&	6.19	&	184	&	284	&	6.17	&	184	&	285	&	5.36	&	184	&	286	&	5.48		\\
	&	185	&	283	&	5.73	&	185	&	284	&	4.79	&	185	&	285	&	4.70	&	185	&	286	&	4.29	&	185	&	287	&	4.29		\\
	&	186	&	284	&	5.43	&	186	&	285	&	4.54	&	186	&	286	&	4.47	&	186	&	287	&	4.02	&	186	&	288	&	3.97		\\
	&	187	&	285	&	4.59	&	187	&	286	&	3.76	&	187	&	287	&	3.60	&	187	&	288	&	3.38	&	187	&	289	&	3.20		\\
	&	188	&	286	&	4.00	&	188	&	287	&	3.30	&	188	&	288	&	3.09	&	188	&	289	&	3.00	&	188	&	290	&	2.73		\\
	&	189	&	287	&	4.13	&	189	&	288	&	3.30	&	189	&	289	&	3.10	&	189	&	290	&	2.88	&	189	&	291	&	2.48		\\
	&	190	&	288	&	3.52	&	190	&	289	&	2.79	&	190	&	290	&	2.48	&	190	&	291	&	2.42	&	190	&	292	&	2.09		\\
	&	191	&	289	&	3.53	&	191	&	290	&	2.69	&	191	&	291	&	2.57	&	191	&	292	&	2.64	&	191	&	293	&	2.24		\\
	&	192	&	290	&	3.08	&	192	&	291	&	2.17	&	192	&	292	&	2.05	&	192	&	293	&	2.13	&	192	&	294	&	1.73		\\

\noalign{\smallskip} \noalign{\smallskip}

\end{tabular}

\end{ruledtabular}

\end{table*}


\endgroup

\begingroup

\squeezetable


\begin{table*}

\caption{\label{4} Calculated fission barrier heights (in MeV).}

 \begin{ruledtabular}

\begin{tabular}{|cccc|ccc|ccc|ccc|ccc|}

&     N &   A  &  $B_{f}$   &      N    &    A  &  $B_{f}$  &   N       &   A   &  $B_{f}$  &        N  &    A  &  $B_{f}$  &        N  &   A   &  $B_{f}$ \\

\noalign{\smallskip}\hline\noalign{\smallskip}


 &       & \textbf{Z=103}&    &           & \textbf{Z=104}&   &           &  \textbf{Z=105}&  &          &  \textbf{Z=106}&  &           & \textbf{Z=107}&   \\

&	139	&	242	&	3.13	&		&		&	    	&		&		&	    	&		&		&	    	&		&		&		\\					
&	140	&	243	&	3.14	&	140	&	244	&	2.69	&		&		&	    	&		&		&	    	&		&		&		\\					
&	141	&	244	&	4.01	&	141	&	245	&	3.58	&	141	&	246	&	2.89	&		&		&	    	&		&		&		\\					
&	142	&	245	&	3.92	&	142	&	246	&	3.46	&	142	&	247	&	3.09	&	142	&	248	&	2.66	&		&		&		\\					
&	143	&	246	&	4.55	&	143	&	247	&	4.16	&	143	&	248	&	3.77	&	143	&	249	&	3.37	&	143	&	250	&	3.30	\\					
&	144	&	247	&	4.66	&	144	&	248	&	4.13	&	144	&	249	&	3.75	&	144	&	250	&	3.35	&	144	&	251	&	3.31	\\					
&	145	&	248	&	5.29	&	145	&	249	&	4.95	&	145	&	250	&	4.68	&	145	&	251	&	4.21	&	145	&	252	&	4.17	\\					
&	146	&	249	&	5.19	&	146	&	250	&	4.74	&	146	&	251	&	4.56	&	146	&	252	&	4.15	&	146	&	253	&	4.16	\\					
&	147	&	250	&	6.10	&	147	&	251	&	5.47	&	147	&	252	&	5.57	&	147	&	253	&	5.01	&	147	&	254	&	5.19	\\					
&	148	&	251	&	5.79	&	148	&	252	&	5.36	&	148	&	253	&	5.62	&	148	&	254	&	4.98	&	148	&	255	&	5.08	\\					
&	149	&	252	&	6.54	&	149	&	253	&	6.16	&	149	&	254	&	6.77	&	149	&	255	&	5.98	&	149	&	256	&	6.22	\\					
&	150	&	253	&	6.28	&	150	&	254	&	5.93	&	150	&	255	&	6.43	&	150	&	256	&	5.76	&	150	&	257	&	5.92	\\					
&	151	&	254	&	7.26	&	151	&	255	&	6.93	&	151	&	256	&	7.50	&	151	&	257	&	6.85	&	151	&	258	&	7.17	\\					
&	152	&	255	&	6.81	&	152	&	256	&	6.44	&	152	&	257	&	7.04	&	152	&	258	&	6.37	&	152	&	259	&	6.70	\\					
&	153	&	256	&	6.62	&	153	&	257	&	6.36	&	153	&	258	&	7.18	&	153	&	259	&	6.58	&	153	&	260	&	6.83	\\					
&	154	&	257	&	6.45	&	154	&	258	&	6.11	&	154	&	259	&	6.99	&	154	&	260	&	6.49	&	154	&	261	&	6.70	\\					
&	155	&	258	&	6.49	&	155	&	259	&	6.14	&	155	&	260	&	7.18	&	155	&	261	&	6.61	&	155	&	262	&	6.69	\\					
&	156	&	259	&	6.30	&	156	&	260	&	5.96	&	156	&	261	&	6.82	&	156	&	262	&	6.30	&	156	&	263	&	6.53	\\					
&	157	&	260	&	6.33	&	157	&	261	&	6.01	&	157	&	262	&	6.84	&	157	&	263	&	6.37	&	157	&	264	&	6.74	\\					
&	158	&	261	&	6.10	&	158	&	262	&	5.73	&	158	&	263	&	6.56	&	158	&	264	&	6.03	&	158	&	265	&	6.44	\\					
&	159	&	262	&	6.15	&	159	&	263	&	5.68	&	159	&	264	&	6.53	&	159	&	265	&	5.97	&	159	&	266	&	6.92	\\					
&	160	&	263	&	5.89	&	160	&	264	&	5.45	&	160	&	265	&	6.31	&	160	&	266	&	5.83	&	160	&	267	&	6.72	\\					
&	161	&	264	&	6.36	&	161	&	265	&	5.91	&	161	&	266	&	6.83	&	161	&	267	&	6.45	&	161	&	268	&	7.60	\\					
&	162	&	265	&	5.83	&	162	&	266	&	5.46	&	162	&	267	&	6.40	&	162	&	268	&	5.95	&	162	&	269	&	7.20	\\					
&	163	&	266	&	5.19	&	163	&	267	&	5.14	&	163	&	268	&	5.99	&	163	&	269	&	5.69	&	163	&	270	&	6.95	\\					
&	164	&	267	&	4.83	&	164	&	268	&	4.59	&	164	&	269	&	5.37	&	164	&	270	&	5.06	&	164	&	271	&	6.27	\\					
&	165	&	268	&	4.17	&	165	&	269	&	4.13	&	165	&	270	&	4.71	&	165	&	271	&	4.50	&	165	&	272	&	5.61	\\					
&	166	&	269	&	3.97	&	166	&	270	&	3.85	&	166	&	271	&	4.43	&	166	&	272	&	4.17	&	166	&	273	&	5.30	\\					
&	167	&	270	&	3.82	&	167	&	271	&	3.67	&	167	&	272	&	4.00	&	167	&	273	&	3.85	&	167	&	274	&	4.67	\\					
&	168	&	271	&	3.41	&	168	&	272	&	3.33	&	168	&	273	&	3.70	&	168	&	274	&	3.54	&	168	&	275	&	4.38	\\					
&	169	&	272	&	3.52	&	169	&	273	&	3.44	&	169	&	274	&	3.67	&	169	&	275	&	3.44	&	169	&	276	&	3.93	\\					
&	170	&	273	&	3.19	&	170	&	274	&	3.12	&	170	&	275	&	3.37	&	170	&	276	&	3.20	&	170	&	277	&	3.72	\\					
&	171	&	274	&	3.68	&	171	&	275	&	3.65	&	171	&	276	&	3.81	&	171	&	277	&	3.56	&	171	&	278	&	4.11	\\					
&	172	&	275	&	3.30	&	172	&	276	&	3.20	&	172	&	277	&	3.36	&	172	&	278	&	3.24	&	172	&	279	&	3.70	\\					
&	173	&	276	&	4.07	&	173	&	277	&	3.83	&	173	&	278	&	4.08	&	173	&	279	&	3.89	&	173	&	280	&	4.33	\\					
&	174	&	277	&	3.67	&	174	&	278	&	3.48	&	174	&	279	&	3.72	&	174	&	280	&	3.55	&	174	&	281	&	4.10	\\					
&	175	&	278	&	4.55	&	175	&	279	&	4.55	&	175	&	280	&	4.73	&	175	&	281	&	4.71	&	175	&	282	&	5.15	\\					
&	176	&	279	&	4.00	&	176	&	280	&	4.12	&	176	&	281	&	4.10	&	176	&	282	&	4.15	&	176	&	283	&	4.64	\\					
&	177	&	280	&	4.97	&	177	&	281	&	5.24	&	177	&	282	&	5.03	&	177	&	283	&	5.29	&	177	&	284	&	5.27	\\					
&	178	&	281	&	4.46	&	178	&	282	&	4.77	&	178	&	283	&	4.51	&	178	&	284	&	4.78	&	178	&	285	&	4.65	\\					
&	179	&	282	&	5.01	&	179	&	283	&	5.19	&	179	&	284	&	5.23	&	179	&	285	&	5.26	&	179	&	286	&	5.49	\\					
&	180	&	283	&	4.68	&	180	&	284	&	4.89	&	180	&	285	&	4.75	&	180	&	286	&	5.06	&	180	&	287	&	4.99	\\					
&	181	&	284	&	5.62	&	181	&	285	&	5.69	&	181	&	286	&	5.71	&	181	&	287	&	5.83	&	181	&	288	&	5.84	\\					
&	182	&	285	&	5.00	&	182	&	286	&	5.17	&	182	&	287	&	5.13	&	182	&	288	&	5.22	&	182	&	289	&	5.27	\\					
&	183	&	286	&	5.91	&	183	&	287	&	5.89	&	183	&	288	&	5.87	&	183	&	289	&	6.01	&	183	&	290	&	6.13	\\					
&	184	&	287	&	5.31	&	184	&	288	&	5.36	&	184	&	289	&	5.34	&	184	&	290	&	5.41	&	184	&	291	&	5.52	\\					
&	185	&	288	&	4.38	&	185	&	289	&	4.27	&	185	&	290	&	4.32	&	185	&	291	&	4.26	&	185	&	292	&	4.25	\\					
&	186	&	289	&	4.17	&	186	&	290	&	4.01	&	186	&	291	&	4.11	&	186	&	292	&	4.11	&	186	&	293	&	4.26	\\					
&	187	&	290	&	2.74	&	187	&	291	&	3.05	&	187	&	292	&	3.24	&	187	&	293	&	2.99	&	187	&	294	&	3.19	\\					
&	188	&	291	&	2.76	&	188	&	292	&	2.57	&	188	&	293	&	2.83	&	188	&	294	&	2.67	&	188	&	295	&	2.91	\\					
&	189	&	292	&	2.44	&	189	&	293	&	2.23	&	189	&	294	&	2.18	&	189	&	295	&	1.88	&	189	&	296	&	1.86	\\					
&	190	&	293	&	2.20	&	190	&	294	&	1.81	&	190	&	295	&	1.92	&	190	&	296	&	1.56	&	190	&	297	&	1.61	\\					
&	191	&	294	&	2.21	&	191	&	295	&	1.89	&	191	&	296	&	1.73	&	191	&	297	&	1.48	&	191	&	298	&	1.40	\\					
&	192	&	295	&	1.70	&	192	&	296	&	1.44	&	192	&	297	&	1.37	&	192	&	298	&	1.12	&	192	&	299	&	1.06	\\

\noalign{\smallskip} \noalign{\smallskip}

\end{tabular}

\end{ruledtabular}

\end{table*}


\endgroup

\begingroup

\squeezetable


\begin{table*}

\caption{\label{4} Calculated fission barrier heights (in MeV).}

 \begin{ruledtabular}

\begin{tabular}{|cccc|ccc|ccc|ccc|ccc|}

&     N &   A  &  $B_{f}$   &      N    &    A  &  $B_{f}$  &   N       &   A   &  $B_{f}$  &        N  &    A  &  $B_{f}$  &        N  &   A   &  $B_{f}$ \\

\noalign{\smallskip}\hline\noalign{\smallskip}


 &       & \textbf{Z=108}&    &           & \textbf{Z=109}&   &           &  \textbf{Z=110}&  &          &  \textbf{Z=111}&  &           & \textbf{Z=112}&   \\

&	144	&	252	&	2.72	&		&		&		    &		&		&		   &		&		&	     	&		&		&		    \\
&	145	&	253	&	3.49	&	145	&	254	&	2.72	&		&		&		   &		&		&		    &		&		&		    \\
&	146	&	254	&	3.58	&	146	&	255	&	3.06	&	146	&	256	&	2.47	&		&		&		    &		&		&		    \\
&	147	&	255	&	4.50	&	147	&	256	&	3.71	&	147	&	257	&	3.29	&	147	&	258	&	2.60	&		&		&		    \\
&	148	&	256	&	4.48	&	148	&	257	&	3.70	&	148	&	258	&	3.32	&	148	&	259	&	2.50	&	148	&	260	&	2.21	\\
&	149	&	257	&	5.60	&	149	&	258	&	4.88	&	149	&	259	&	4.26	&	149	&	260	&	3.33	&	149	&	261	&	2.96	\\
&	150	&	258	&	5.29	&	150	&	259	&	4.53	&	150	&	260	&	4.05	&	150	&	261	&	3.27	&	150	&	262	&	2.77	\\
&	151	&	259	&	6.40	&	151	&	260	&	5.53	&	151	&	261	&	5.12	&	151	&	262	&	4.10	&	151	&	263	&	3.92	\\
&	152	&	260	&	5.98	&	152	&	261	&	5.17	&	152	&	262	&	4.68	&	152	&	263	&	3.67	&	152	&	264	&	3.47	\\
&	153	&	261	&	6.13	&	153	&	262	&	5.00	&	153	&	263	&	4.68	&	153	&	264	&	3.73	&	153	&	265	&	3.40	\\
&	154	&	262	&	6.07	&	154	&	263	&	4.97	&	154	&	264	&	4.50	&	154	&	265	&	3.81	&	154	&	266	&	3.34	\\
&	155	&	263	&	6.07	&	155	&	264	&	5.01	&	155	&	265	&	4.50	&	155	&	266	&	3.65	&	155	&	267	&	3.22	\\
&	156	&	264	&	5.93	&	156	&	265	&	4.92	&	156	&	266	&	4.39	&	156	&	267	&	3.74	&	156	&	268	&	3.26	\\
&	157	&	265	&	6.09	&	157	&	266	&	5.29	&	157	&	267	&	4.50	&	157	&	268	&	3.65	&	157	&	269	&	3.21	\\
&	158	&	266	&	5.83	&	158	&	267	&	5.11	&	158	&	268	&	4.38	&	158	&	269	&	3.68	&	158	&	270	&	3.31	\\
&	159	&	267	&	6.19	&	159	&	268	&	5.85	&	159	&	269	&	5.04	&	159	&	270	&	4.52	&	159	&	271	&	3.98	\\
&	160	&	268	&	5.93	&	160	&	269	&	5.65	&	160	&	270	&	4.86	&	160	&	271	&	4.26	&	160	&	272	&	3.78	\\
&	161	&	269	&	6.87	&	161	&	270	&	6.72	&	161	&	271	&	5.91	&	161	&	272	&	5.58	&	161	&	273	&	5.02	\\
&	162	&	270	&	6.46	&	162	&	271	&	6.44	&	162	&	272	&	5.62	&	162	&	273	&	5.30	&	162	&	274	&	4.72	\\
&	163	&	271	&	6.28	&	163	&	272	&	6.48	&	163	&	273	&	5.85	&	163	&	274	&	5.70	&	163	&	275	&	5.00	\\
&	164	&	272	&	5.52	&	164	&	273	&	5.85	&	164	&	274	&	5.22	&	164	&	275	&	5.03	&	164	&	276	&	4.46	\\
&	165	&	273	&	5.00	&	165	&	274	&	5.56	&	165	&	275	&	4.87	&	165	&	276	&	5.05	&	165	&	277	&	4.46	\\
&	166	&	274	&	4.62	&	166	&	275	&	5.14	&	166	&	276	&	4.47	&	166	&	277	&	4.62	&	166	&	278	&	4.01	\\
&	167	&	275	&	4.16	&	167	&	276	&	4.90	&	167	&	277	&	4.16	&	167	&	278	&	4.61	&	167	&	279	&	3.99	\\
&	168	&	276	&	3.80	&	168	&	277	&	4.45	&	168	&	278	&	3.73	&	168	&	279	&	4.13	&	168	&	280	&	3.78	\\
&	169	&	277	&	3.40	&	169	&	278	&	4.14	&	169	&	279	&	3.44	&	169	&	280	&	4.35	&	169	&	281	&	3.88	\\
&	170	&	278	&	3.20	&	170	&	279	&	3.84	&	170	&	280	&	3.29	&	170	&	281	&	4.19	&	170	&	282	&	3.74	\\
&	171	&	279	&	3.80	&	171	&	280	&	4.09	&	171	&	281	&	3.95	&	171	&	282	&	4.74	&	171	&	283	&	4.58	\\
&	172	&	280	&	3.31	&	172	&	281	&	3.87	&	172	&	282	&	3.72	&	172	&	283	&	4.67	&	172	&	284	&	4.34	\\
&	173	&	281	&	4.20	&	173	&	282	&	4.55	&	173	&	283	&	4.68	&	173	&	284	&	5.32	&	173	&	285	&	5.26	\\
&	174	&	282	&	3.88	&	174	&	283	&	4.22	&	174	&	284	&	4.40	&	174	&	285	&	5.07	&	174	&	286	&	5.03	\\
&	175	&	283	&	5.00	&	175	&	284	&	5.37	&	175	&	285	&	5.51	&	175	&	286	&	5.83	&	175	&	287	&	6.04	\\
&	176	&	284	&	4.50	&	176	&	285	&	4.91	&	176	&	286	&	5.04	&	176	&	287	&	5.37	&	176	&	288	&	5.60	\\
&	177	&	285	&	5.43	&	177	&	286	&	5.74	&	177	&	287	&	5.86	&	177	&	288	&	6.39	&	177	&	289	&	6.29	\\
&	178	&	286	&	4.93	&	178	&	287	&	5.17	&	178	&	288	&	5.33	&	178	&	289	&	5.93	&	178	&	290	&	5.87	\\
&	179	&	287	&	5.61	&	179	&	288	&	6.03	&	179	&	289	&	5.96	&	179	&	290	&	6.60	&	179	&	291	&	6.32	\\
&	180	&	288	&	5.29	&	180	&	289	&	5.66	&	180	&	290	&	5.61	&	180	&	291	&	6.23	&	180	&	292	&	5.94	\\
&	181	&	289	&	6.06	&	181	&	290	&	6.19	&	181	&	291	&	6.12	&	181	&	292	&	6.59	&	181	&	293	&	6.29	\\
&	182	&	290	&	5.47	&	182	&	291	&	5.70	&	182	&	292	&	5.63	&	182	&	293	&	6.18	&	182	&	294	&	5.89	\\
&	183	&	291	&	6.05	&	183	&	292	&	6.46	&	183	&	293	&	6.20	&	183	&	294	&	6.75	&	183	&	295	&	6.48	\\
&	184	&	292	&	5.61	&	184	&	293	&	5.95	&	184	&	294	&	5.68	&	184	&	295	&	6.20	&	184	&	296	&	5.91	\\
&	185	&	293	&	4.20	&	185	&	294	&	4.51	&	185	&	295	&	4.38	&	185	&	296	&	4.79	&	185	&	297	&	4.54	\\
&	186	&	294	&	4.23	&	186	&	295	&	4.68	&	186	&	296	&	4.50	&	186	&	297	&	5.04	&	186	&	298	&	4.74	\\
&	187	&	295	&	3.04	&	187	&	296	&	3.09	&	187	&	297	&	3.07	&	187	&	298	&	3.73	&	187	&	299	&	3.39	\\
&	188	&	296	&	2.83	&	188	&	297	&	3.11	&	188	&	298	&	3.01	&	188	&	299	&	3.76	&	188	&	300	&	3.44	\\
&	189	&	297	&	1.75	&	189	&	298	&	1.80	&	189	&	299	&	1.79	&	189	&	300	&	2.20	&	189	&	301	&	1.92	\\
&	190	&	298	&	1.44	&	190	&	299	&	1.74	&	190	&	300	&	1.58	&	190	&	301	&	2.28	&	190	&	302	&	2.02	\\
&	191	&	299	&	1.30	&	191	&	300	&	1.21	&	191	&	301	&	1.28	&	191	&	302	&	1.05	&	191	&	303	&	1.07	\\
&	192	&	300	&	0.75	&	192	&	301	&	0.74	&	192	&	302	&	0.76	&	192	&	303	&	0.99	&	192	&	304	&	0.76	\\

\noalign{\smallskip} \noalign{\smallskip}

\end{tabular}

\end{ruledtabular}

\end{table*}


\endgroup

\begingroup

\squeezetable


\begin{table*}

\caption{\label{4} Calculated fission barrier heights (in MeV).}

 \begin{ruledtabular}

\begin{tabular}{|cccc|ccc|ccc|ccc|ccc|}

&     N &   A  &  $B_{f}$   &      N    &    A  &  $B_{f}$  &   N       &   A   &  $B_{f}$  &        N  &    A  &  $B_{f}$  &        N  &   A   &  $B_{f}$ \\

\noalign{\smallskip}\hline\noalign{\smallskip}


 &       & \textbf{Z=113}&    &           & \textbf{Z=114}&   &           &  \textbf{Z=115}&  &          &  \textbf{Z=116}&  &           & \textbf{Z=117}&   \\
																								
&	149	&	262	&	2.02	&		&		&		    &		&		&	      	&		&		&	     	&		&		&		\\				
&	150	&	263	&	2.02	&	150	&	264	&	1.72	&		&		&	     	&		&		&	    	&		&		&		\\				
&	151	&	264	&	2.95	&	151	&	265	&	2.75	&	151	&	266	&	1.54	&		&		&		    &		&		&		\\				
&	152	&	265	&	2.44	&	152	&	266	&	2.27	&	152	&	267	&	1.17	&	152	&	268	&	1.08	&		&		&		\\				
&	153	&	266	&	2.33	&	153	&	267	&	2.15	&	153	&	268	&	0.90	&	153	&	269	&	0.96	&	153	&	270	&	0.76	\\				
&	154	&	267	&	2.25	&	154	&	268	&	2.14	&	154	&	269	&	1.34	&	154	&	270	&	1.01	&	154	&	271	&	0.87	\\				
&	155	&	268	&	2.44	&	155	&	269	&	2.13	&	155	&	270	&	2.14	&	155	&	271	&	1.39	&	155	&	272	&	1.93	\\				
&	156	&	269	&	2.34	&	156	&	270	&	2.27	&	156	&	271	&	2.29	&	156	&	272	&	1.72	&	156	&	273	&	1.99	\\				
&	157	&	270	&	2.55	&	157	&	271	&	2.36	&	157	&	272	&	2.81	&	157	&	273	&	2.24	&	157	&	274	&	2.58	\\				
&	158	&	271	&	2.66	&	158	&	272	&	2.50	&	158	&	273	&	2.89	&	158	&	274	&	2.37	&	158	&	275	&	2.91	\\				
&	159	&	272	&	3.65	&	159	&	273	&	3.25	&	159	&	274	&	3.67	&	159	&	275	&	3.13	&	159	&	276	&	3.33	\\				
&	160	&	273	&	3.45	&	160	&	274	&	3.39	&	160	&	275	&	3.81	&	160	&	276	&	3.27	&	160	&	277	&	3.69	\\				
&	161	&	274	&	4.58	&	161	&	275	&	4.53	&	161	&	276	&	4.88	&	161	&	277	&	4.45	&	161	&	278	&	4.74	\\				
&	162	&	275	&	4.41	&	162	&	276	&	4.35	&	162	&	277	&	4.68	&	162	&	278	&	4.18	&	162	&	279	&	4.35	\\				
&	163	&	276	&	4.88	&	163	&	277	&	4.75	&	163	&	278	&	5.30	&	163	&	279	&	4.77	&	163	&	280	&	5.10	\\				
&	164	&	277	&	4.44	&	164	&	278	&	4.32	&	164	&	279	&	4.79	&	164	&	280	&	4.34	&	164	&	281	&	4.66	\\				
&	165	&	278	&	4.58	&	165	&	279	&	4.29	&	165	&	280	&	4.87	&	165	&	281	&	4.36	&	165	&	282	&	4.47	\\				
&	166	&	279	&	4.23	&	166	&	280	&	3.99	&	166	&	281	&	4.56	&	166	&	282	&	4.16	&	166	&	283	&	4.53	\\				
&	167	&	280	&	4.67	&	167	&	281	&	4.37	&	167	&	282	&	4.96	&	167	&	283	&	4.52	&	167	&	284	&	4.77	\\				
&	168	&	281	&	4.34	&	168	&	282	&	4.05	&	168	&	283	&	4.81	&	168	&	284	&	4.48	&	168	&	285	&	4.69	\\				
&	169	&	282	&	4.89	&	169	&	283	&	4.52	&	169	&	284	&	5.29	&	169	&	285	&	4.83	&	169	&	286	&	5.23	\\				
&	170	&	283	&	4.46	&	170	&	284	&	4.48	&	170	&	285	&	4.98	&	170	&	286	&	4.70	&	170	&	287	&	5.30	\\				
&	171	&	284	&	5.19	&	171	&	285	&	5.16	&	171	&	286	&	5.70	&	171	&	287	&	5.76	&	171	&	288	&	6.21	\\				
&	172	&	285	&	4.98	&	172	&	286	&	4.83	&	172	&	287	&	5.56	&	172	&	288	&	5.45	&	172	&	289	&	5.95	\\				
&	173	&	286	&	5.74	&	173	&	287	&	5.76	&	173	&	288	&	6.21	&	173	&	289	&	6.18	&	173	&	290	&	6.81	\\				
&	174	&	287	&	5.54	&	174	&	288	&	5.52	&	174	&	289	&	6.02	&	174	&	290	&	5.93	&	174	&	291	&	6.46	\\				
&	175	&	288	&	6.43	&	175	&	289	&	6.43	&	175	&	290	&	6.68	&	175	&	291	&	6.67	&	175	&	292	&	7.04	\\				
&	176	&	289	&	6.28	&	176	&	290	&	6.04	&	176	&	291	&	6.55	&	176	&	292	&	6.31	&	176	&	293	&	6.56	\\				
&	177	&	290	&	6.95	&	177	&	291	&	6.67	&	177	&	292	&	7.46	&	177	&	293	&	6.82	&	177	&	294	&	7.31	\\				
&	178	&	291	&	6.61	&	178	&	292	&	6.58	&	178	&	293	&	6.80	&	178	&	294	&	6.37	&	178	&	295	&	6.64	\\				
&	179	&	292	&	7.26	&	179	&	293	&	6.75	&	179	&	294	&	7.08	&	179	&	295	&	6.64	&	179	&	296	&	6.88	\\				
&	180	&	293	&	6.82	&	180	&	294	&	6.45	&	180	&	295	&	6.69	&	180	&	296	&	6.25	&	180	&	297	&	6.32	\\				
&	181	&	294	&	6.93	&	181	&	295	&	6.64	&	181	&	296	&	7.00	&	181	&	297	&	6.51	&	181	&	298	&	6.70	\\				
&	182	&	295	&	6.71	&	182	&	296	&	6.29	&	182	&	297	&	6.48	&	182	&	298	&	6.10	&	182	&	299	&	6.12	\\				
&	183	&	296	&	7.13	&	183	&	297	&	6.58	&	183	&	298	&	6.75	&	183	&	299	&	6.35	&	183	&	300	&	6.32	\\				
&	184	&	297	&	6.63	&	184	&	298	&	6.10	&	184	&	299	&	6.23	&	184	&	300	&	5.84	&	184	&	301	&	5.81	\\				
&	185	&	298	&	5.36	&	185	&	299	&	4.72	&	185	&	300	&	4.97	&	185	&	301	&	4.39	&	185	&	302	&	4.27	\\				
&	186	&	299	&	5.43	&	186	&	300	&	4.92	&	186	&	301	&	5.01	&	186	&	302	&	4.64	&	186	&	303	&	4.61	\\				
&	187	&	300	&	4.11	&	187	&	301	&	3.52	&	187	&	302	&	3.85	&	187	&	303	&	3.27	&	187	&	304	&	3.32	\\				
&	188	&	301	&	4.14	&	188	&	302	&	3.66	&	188	&	303	&	3.74	&	188	&	304	&	3.40	&	188	&	305	&	3.34	\\				
&	189	&	302	&	2.67	&	189	&	303	&	2.26	&	189	&	304	&	2.53	&	189	&	305	&	2.04	&	189	&	306	&	2.28	\\				
&	190	&	303	&	2.70	&	190	&	304	&	2.36	&	190	&	305	&	2.34	&	190	&	306	&	2.16	&	190	&	307	&	2.07	\\				
&	191	&	304	&	1.30	&	191	&	305	&	0.78	&	191	&	306	&	1.53	&	191	&	307	&	0.90	&	191	&	308	&	1.25	\\				
&	192	&	305	&	1.28	&	192	&	306	&	0.91	&	192	&	307	&	1.23	&	192	&	308	&	0.75	&	192	&	309	&	1.18	\\

\noalign{\smallskip} \noalign{\smallskip}

\end{tabular}

\end{ruledtabular}

\end{table*}


\endgroup

\begingroup

\squeezetable


\begin{table*}

\caption{\label{4} Calculated fission barrier heights (in MeV).}

 \begin{ruledtabular}

\begin{tabular}{|cccc|ccc|ccc|ccc|ccc|}

&     N &   A  &  $B_{f}$   &      N    &    A  &  $B_{f}$  &   N       &   A   &  $B_{f}$  &        N  &    A  &  $B_{f}$  &        N  &   A   &  $B_{f}$ \\

\noalign{\smallskip}\hline\noalign{\smallskip}


 &       & \textbf{Z=118}&    &           & \textbf{Z=119}&   &           &  \textbf{Z=120}&  &          &  \textbf{Z=121}&  &           & \textbf{Z=122}&   \\
																								
&	154	&	272	&	0.66	&		&		&               &		&		&		&		&		&	    	&		&		&		\\				
&	155	&	273	&	1.39	&	155	&	274	&	1.82	&		&		&       	&		&		&	     	&		&		&		\\				
&	156	&	274	&	1.41	&	156	&	275	&	1.83	&	156	&	276	&	1.38	&		&		&	     	&		&		&		\\				
&	157	&	275	&	2.30	&	157	&	276	&	2.26	&	157	&	277	&	1.79	&	157	&	278	&	1.96	&		&		&		\\				
&	158	&	276	&	2.25	&	158	&	277	&	2.25	&	158	&	278	&	1.88	&	158	&	279	&	2.11	&	158	&	280	&	1.47	\\				
&	159	&	277	&	3.06	&	159	&	278	&	2.78	&	159	&	279	&	2.39	&	159	&	280	&	2.72	&	159	&	281	&	1.94	\\				
&	160	&	278	&	3.15	&	160	&	279	&	2.79	&	160	&	280	&	2.44	&	160	&	281	&	2.75	&	160	&	282	&	2.15	\\				
&	161	&	279	&	4.30	&	161	&	280	&	3.49	&	161	&	281	&	3.23	&	161	&	282	&	3.66	&	161	&	283	&	3.07	\\				
&	162	&	280	&	3.94	&	162	&	281	&	3.27	&	162	&	282	&	3.07	&	162	&	283	&	3.50	&	162	&	284	&	2.94	\\				
&	163	&	281	&	4.32	&	163	&	282	&	3.96	&	163	&	283	&	3.89	&	163	&	284	&	4.56	&	163	&	285	&	3.88	\\				
&	164	&	282	&	4.13	&	164	&	283	&	3.68	&	164	&	284	&	3.57	&	164	&	285	&	4.16	&	164	&	286	&	3.61	\\				
&	165	&	283	&	3.79	&	165	&	284	&	3.87	&	165	&	285	&	3.65	&	165	&	286	&	4.67	&	165	&	287	&	4.46	\\				
&	166	&	284	&	3.64	&	166	&	285	&	3.67	&	166	&	286	&	3.43	&	166	&	287	&	4.36	&	166	&	288	&	3.91	\\				
&	167	&	285	&	4.09	&	167	&	286	&	4.43	&	167	&	287	&	4.08	&	167	&	288	&	4.40	&	167	&	289	&	4.06	\\				
&	168	&	286	&	4.01	&	168	&	287	&	4.45	&	168	&	288	&	4.21	&	168	&	289	&	4.37	&	168	&	290	&	4.01	\\				
&	169	&	287	&	5.03	&	169	&	288	&	5.21	&	169	&	289	&	5.01	&	169	&	290	&	5.35	&	169	&	291	&	4.83	\\				
&	170	&	288	&	5.05	&	170	&	289	&	5.23	&	170	&	290	&	5.01	&	170	&	291	&	5.41	&	170	&	292	&	4.80	\\				
&	171	&	289	&	6.03	&	171	&	290	&	6.26	&	171	&	291	&	6.08	&	171	&	292	&	6.56	&	171	&	293	&	5.96	\\				
&	172	&	290	&	5.75	&	172	&	291	&	5.75	&	172	&	292	&	5.48	&	172	&	293	&	5.90	&	172	&	294	&	5.36	\\				
&	173	&	291	&	6.40	&	173	&	292	&	6.55	&	173	&	293	&	6.06	&	173	&	294	&	6.15	&	173	&	295	&	5.54	\\				
&	174	&	292	&	6.09	&	174	&	293	&	6.21	&	174	&	294	&	5.62	&	174	&	295	&	5.89	&	174	&	296	&	5.28	\\				
&	175	&	293	&	6.62	&	175	&	294	&	6.95	&	175	&	295	&	6.28	&	175	&	296	&	6.00	&	175	&	297	&	5.48	\\				
&	176	&	294	&	6.09	&	176	&	295	&	6.32	&	176	&	296	&	5.79	&	176	&	297	&	5.69	&	176	&	298	&	5.10	\\				
&	177	&	295	&	6.64	&	177	&	296	&	6.71	&	177	&	297	&	6.02	&	177	&	298	&	5.90	&	177	&	299	&	5.36	\\				
&	178	&	296	&	6.12	&	178	&	297	&	6.20	&	178	&	298	&	5.56	&	178	&	299	&	5.38	&	178	&	300	&	4.86	\\				
&	179	&	297	&	6.21	&	179	&	298	&	6.20	&	179	&	299	&	5.57	&	179	&	300	&	5.38	&	179	&	301	&	4.75	\\				
&	180	&	298	&	5.79	&	180	&	299	&	5.72	&	180	&	300	&	5.08	&	180	&	301	&	4.82	&	180	&	302	&	4.26	\\				
&	181	&	299	&	6.02	&	181	&	300	&	5.88	&	181	&	301	&	5.24	&	181	&	302	&	4.99	&	181	&	303	&	4.43	\\				
&	182	&	300	&	5.52	&	182	&	301	&	5.38	&	182	&	302	&	4.71	&	182	&	303	&	4.37	&	182	&	304	&	3.78	\\				
&	183	&	301	&	5.71	&	183	&	302	&	5.50	&	183	&	303	&	4.81	&	183	&	304	&	4.27	&	183	&	305	&	3.70	\\				
&	184	&	302	&	5.20	&	184	&	303	&	4.98	&	184	&	304	&	4.30	&	184	&	305	&	3.64	&	184	&	306	&	3.16	\\				
&	185	&	303	&	3.93	&	185	&	304	&	3.82	&	185	&	305	&	3.02	&	185	&	306	&	2.96	&	185	&	307	&	2.27	\\				
&	186	&	304	&	4.03	&	186	&	305	&	3.85	&	186	&	306	&	3.14	&	186	&	307	&	2.60	&	186	&	308	&	2.08	\\				
&	187	&	305	&	2.80	&	187	&	306	&	2.78	&	187	&	307	&	2.17	&	187	&	308	&	2.15	&	187	&	309	&	1.66	\\				
&	188	&	306	&	2.78	&	188	&	307	&	2.63	&	188	&	308	&	1.95	&	188	&	309	&	1.84	&	188	&	310	&	1.50	\\				
&	189	&	307	&	1.78	&	189	&	308	&	2.03	&	189	&	309	&	1.51	&	189	&	310	&	1.50	&	189	&	311	&	1.31	\\				
&	190	&	308	&	1.57	&	190	&	309	&	1.97	&	190	&	310	&	1.39	&	190	&	311	&	1.45	&	190	&	312	&	1.15	\\				
&	191	&	309	&	0.79	&	191	&	310	&	1.67	&	191	&	311	&	1.05	&	191	&	312	&	1.20	&	191	&	313	&	0.68	\\				
&	192	&	310	&	0.80	&	192	&	311	&	1.63	&	192	&	312	&	1.05	&	192	&	313	&	1.28	&	192	&	314	&	0.80	\\				

\noalign{\smallskip} \noalign{\smallskip}

\end{tabular}

\end{ruledtabular}

\end{table*}


\endgroup

\begingroup

\squeezetable


\begin{table*}

\caption{\label{4} Calculated fission barrier heights (in MeV).}

 \begin{ruledtabular}

\begin{tabular}{|cccc|ccc|ccc|ccc|}

&     N &   A  &  $B_{f}$   &      N    &    A  &  $B_{f}$  &   N       &   A   &  $B_{f}$  &        N  &    A  &  $B_{f}$   \\

\noalign{\smallskip}\hline\noalign{\smallskip}


 &       & \textbf{Z=123}&    &           & \textbf{Z=124}&   &           &  \textbf{Z=125}&  &          &  \textbf{Z=126}&    \\
					    																			
&	159	&	282	&	2.14	&		&		&		&		&		&	    	&		&		&		\\
&	160	&	283	&	2.25	&	160	&	284	&	1.78	&		&		&		&		&		&		\\
&	161	&	284	&	3.01	&	161	&	285	&	2.51	&	161	&	286	&	2.47	&		&		&		\\
&	162	&	285	&	2.87	&	162	&	286	&	2.44	&	162	&	287	&	2.44	&	162	&	288	&	1.89	\\
&	163	&	286	&	4.03	&	163	&	287	&	3.46	&	163	&	288	&	3.18	&	163	&	289	&	2.64	\\
&	164	&	287	&	3.69	&	164	&	288	&	3.24	&	164	&	289	&	3.18	&	164	&	290	&	2.58	\\
&	165	&	288	&	4.75	&	165	&	289	&	4.43	&	165	&	290	&	4.43	&	165	&	291	&	3.85	\\
&	166	&	289	&	4.15	&	166	&	290	&	3.78	&	166	&	291	&	3.76	&	166	&	292	&	3.24	\\
&	167	&	290	&	4.43	&	167	&	291	&	4.08	&	167	&	292	&	4.14	&	167	&	293	&	3.69	\\
&	168	&	291	&	4.08	&	168	&	292	&	3.54	&	168	&	293	&	3.64	&	168	&	294	&	3.15	\\
&	169	&	292	&	5.10	&	169	&	293	&	4.44	&	169	&	294	&	4.34	&	169	&	295	&	3.48	\\
&	170	&	293	&	4.94	&	170	&	294	&	4.38	&	170	&	295	&	4.35	&	170	&	296	&	3.49	\\
&	171	&	294	&	5.96	&	171	&	295	&	5.44	&	171	&	296	&	5.47	&	171	&	297	&	4.62	\\
&	172	&	295	&	5.52	&	172	&	296	&	5.01	&	172	&	297	&	5.06	&	172	&	298	&	4.20	\\
&	173	&	296	&	5.52	&	173	&	297	&	5.10	&	173	&	298	&	5.30	&	173	&	299	&	4.84	\\
&	174	&	297	&	5.23	&	174	&	298	&	4.81	&	174	&	299	&	5.02	&	174	&	300	&	4.52	\\
&	175	&	298	&	5.16	&	175	&	299	&	4.82	&	175	&	300	&	5.13	&	175	&	301	&	4.65	\\
&	176	&	299	&	4.82	&	176	&	300	&	4.51	&	176	&	301	&	4.76	&	176	&	302	&	4.37	\\
&	177	&	300	&	5.26	&	177	&	301	&	4.92	&	177	&	302	&	5.14	&	177	&	303	&	4.75	\\
&	178	&	301	&	4.75	&	178	&	302	&	4.48	&	178	&	303	&	4.66	&	178	&	304	&	4.28	\\
&	179	&	302	&	4.83	&	179	&	303	&	4.55	&	179	&	304	&	4.68	&	179	&	305	&	4.33	\\
&	180	&	303	&	4.07	&	180	&	304	&	3.84	&	180	&	305	&	4.03	&	180	&	306	&	3.66	\\
&	181	&	304	&	3.90	&	181	&	305	&	3.39	&	181	&	306	&	3.45	&	181	&	307	&	2.96	\\
&	182	&	305	&	3.39	&	182	&	306	&	2.95	&	182	&	307	&	3.17	&	182	&	308	&	2.79	\\
&	183	&	306	&	3.54	&	183	&	307	&	2.78	&	183	&	308	&	3.02	&	183	&	309	&	2.21	\\
&	184	&	307	&	2.91	&	184	&	308	&	2.25	&	184	&	309	&	2.53	&	184	&	310	&	2.08	\\
&	185	&	308	&	2.20	&	185	&	309	&	2.07	&	185	&	310	&	2.40	&	185	&	311	&	2.05	\\
&	186	&	309	&	2.01	&	186	&	310	&	1.96	&	186	&	311	&	2.37	&	186	&	312	&	1.90	\\
&	187	&	310	&	1.89	&	187	&	311	&	1.69	&	187	&	312	&	1.90	&	187	&	313	&	1.39	\\
&	188	&	311	&	1.85	&	188	&	312	&	1.62	&	188	&	313	&	1.76	&	188	&	314	&	1.26	\\
&	189	&	312	&	1.47	&	189	&	313	&	1.16	&	189	&	314	&	1.36	&	189	&	315	&	0.90	\\
&	190	&	313	&	1.31	&	190	&	314	&	0.98	&	190	&	315	&	1.10	&	190	&	316	&	0.80	\\
&	191	&	314	&	0.90	&	191	&	315	&	0.68	&	191	&	316	&	0.91	&	191	&	317	&	0.81	\\
&	192	&	315	&	0.76	&	192	&	316	&	0.62	&	192	&	317	&	0.84	&	192	&	318	&	0.70	\\

\noalign{\smallskip} \noalign{\smallskip}

\end{tabular}

\end{ruledtabular}

\end{table*}


\endgroup

\end{document}